\documentclass[journal]{new-aiaa}
\usepackage[utf8]{inputenc}
\usepackage{bm}
\usepackage{graphicx}
\usepackage{amsmath}
\usepackage[version=4]{mhchem}
\usepackage{siunitx}
\usepackage{float}
\usepackage{multirow}
\usepackage{longtable,tabularx}
\title{Diffractive-Sail Single-Impulse Reachable Set for Interplanetary Transfer Design}

\author{Shuyue Fu \footnote{PhD Candidate, School of Astronautics, Shen Yuan Honors College, fushuyue@buaa.edu.cn.}, Jinkai Zhang \footnote{Graduate Student, School of Astronautics, sy2515109zjk@buaa.edu.cn.}, Di Wu \footnote{Associate Professor, School of Astronautics, wudi2025@buaa.edu.cn, Member AIAA.}, Peng Shi\footnote{Professor, School of Astronautics, shipeng@buaa.edu.cn.}, and Shengping Gong \footnote{Professor, School of Astronautics, gongsp@buaa.edu.cn, Senior Member AIAA (Corresponding Author).}}
\affil{Beihang University, Beijing, 100191, People's Republic of China}
\affil{State Key Laboratory of High-Efficiency Reusable Aerospace Transportation Technology, Beijing, 102206, People's Republic of China}

\begin{document}

\maketitle

\begin{abstract}
Interest in planetary exploration has renewed, and the design of interplanetary transfers has attracted remarkable attention. This paper considers the interplanetary transfer design using a diffractive sail. Considering a nonzero departure hyperbolic excess velocity, the interplanetary transfer problem is transformed into the problem of computing single-impulse reachable sets. Then, based on previous work, a complementary computational method for reachable sets under arbitrary dynamics is proposed using differential algebra combined with adaptive grid refinement. The adaptive grid refinement considers two types of merit scores that reveal dynamical properties and the truncation error of the differential algebra propagation. The proposed method is applied to compute the diffractive-sail reachable sets, and the results verify the effectiveness of the method and merit scores. Finally, a preliminary design of the interplanetary transfers, specified as the Earth-Mars transfers, is performed based on the diffractive-sail reachable sets. The design results are presented. The effects of the corresponding parameters, including transfer time, diffractive angle, and type of diffractive sails, on transfer characteristics are analyzed, providing further insight into parameter selection for interplanetary transfer design.
\end{abstract}

\section{Introduction}
\lettrine{I}{nterplanetary} transfers have attracted remarkable attention recently due to the proposal and implementation of several planetary exploration missions, such as \textit{Tianwen-1} \cite{zou2021scientific} and \textit{Tianwen-3} \cite{hou2025search} to explore Mars, and \textit{DAVINCI} \cite{garvin2022revealing} to explore Venus. To perform interplanetary transfers, a change in the semi-major axis of the heliocentric orbit is generally required because the planets have different heliocentric distances. Except for Hohmann transfers and Lambert transfers (direct transfers) only using the impulses (chemical propulsion) \cite{battin1999introduction}, the use of planetary gravity assists \cite{strange2002graphical,cao2024semi,li2026interplanetary} and other propulsion ways, such as electric propulsion or low-thrust propulsion \cite{hofmann2021rapid,wu2022analytical,wu2021warm}, solar-sail propulsion \cite{quarta2022solar,quarta2023solar,quarta2023optimal,chu2024minimum,quarta2026optimal}, can further effectively change the semi-major axis of the heliocentric orbit. Among these ways, planetary gravity assists typically require relatively long transfer times, and low-thrust propulsion requires continuous propellant consumption. In contrast, solar-sail propulsion uses the solar radiation pressure (SRP) to accelerate the spacecraft and does not require propellant consumption. Moreover, interplanetary transfers using a solar sail require relatively short transfer time (for example, when using Venus gravity assist to achieve the Earth-Mars transfers, the transfers approximately require 1 and 1.5 years \cite{okutsu2002mars}; however, when using the solar-sail propulsion, the transfers only require 0.5 and 0.75 years with the lower departure hyperbolic excess velocity, which is presented in Section \ref{sec4} in the following texts). Therefore, this paper considers the interplanetary transfer design using solar-sail propulsion. Furthermore, the solar sail can be categorized into reflective sail \cite{mcinnes2004solar,macdonald2005realistic,quarta2022solar} and diffractive sail \cite{swartzlander2017radiation,chu2024minimum,quarta2026optimal}. Compared with the reflective sail, the diffractive sail, first proposed by  Swartzlander \cite{swartzlander2017radiation}, can utilize the force parallel to the sail’s surface generated by the SRP more effectively and enhance transfer performance in terms of transfer time \cite{quarta2023solar,chu2024minimum}. Therefore, this paper adopts the diffractive sail for designing interplanetary transfers. 

Several scholars have studied interplanetary transfer trajectories using a diffractive sail with optimal control methods \cite{quarta2023optimal,chu2024minimum,quarta2026optimal}. The interplanetary transfer problem is usually transformed into the time-optimal control problem. Quarta and Mengali \cite{quarta2023solar} investigated the transfers to Venus, Mars, and Jupiter. Electro-optically controlled sail film is used to generate control inputs corresponding to the SRP acceleration. Moreover, Quarta et al. \cite{quarta2023optimal} further adopted the sail clock angle as the control variable and investigated the transfer to 16 Psyche. Based on their work, Chu and Gong \cite{chu2024minimum} investigated time-optimal interplanetary transfers under various diffractive angles, also adopting the sail clock angle as the control variable. These works pioneered interplanetary transfer design using a diffractive sail. However, these works assumed zero departure hyperbolic excess velocity. A nonzero departure hyperbolic excess velocity represents the capability of launch vehicle engines and can be effectively utilized in the fast interplanetary transfer \cite{zeng2014fast}. For the diffractive-sail scenario, Quarta \cite{quarta2026optimal} further considered the departure hyperbolic excess velocity and proposed the concept of diffractive-sail augmented Hohmann transfer. He considered the tangential departure hyperbolic excess velocity and investigated the control inputs to minimize the sum of departure and arrival hyperbolic excess velocities. Based on his pioneering work, we aim to extend the diffractive-sail interplanetary trajectory design framework to consider a nonzero departure hyperbolic excess velocity. Neglecting the inclinations of the common planets in the solar system, this paper adopts the planar problem with the solar gravity and the SRP generated by an ideal achromatic first-order grating \cite{chu2024potential}. Differing from Quarta's work \cite{quarta2026optimal}, which considered the tangential departure hyperbolic excess velocity, we aim to consider departure hyperbolic excess velocities with arbitrary directions. For the planar problem, we consider the constant diffractive-angle case during trajectory propagation. Therefore, the diffractive-sail dynamics can be treated as one without solving the control inputs, which is similar to the simple two-body problem or an uncontrolled multi-body problem (e.g., circular restricted three-body problem \cite{Koon2001}). Differing from the simple two-body problem, there are no closed-form solutions in the diffractive-sail dynamics. Therefore, numerical methods, including generation of initial guesses and trajectory correction \cite{topputo2013optimal,oshima2019low}, should be adopted to design interplanetary transfers. The problem of generating initial guesses of diffractive-sail interplanetary transfers to a specific planet (with arbitrary-direction nonzero departure hyperbolic excess velocities) can be further extended into a reachability analysis problem. The reachability analysis problem can be further specified as computing single-impulse reachable sets (here, the impulse denotes the departure hyperbolic excess velocities). Therefore, to effectively generate initial guesses for the interplanetary transfer design and also provide some useful insight into the parameter (which can affect the reachable-set configurations) selection for the interplanetary transfer, this paper proposes the concept of the diffractive-sail single-impulse reachable set, investigates its computational method, and applies it to the interplanetary transfer design.

In astrodynamics, the reachable set usually describes the set of the states of spacecraft with the presented initial states following specific control inputs and dynamics \cite{zhou2025single}. According to ways of the control inputs, the reachable set can be categorized into impulsive (single-impulse and multi-impulse) reachable set \cite{wen2023reachable,zhang2024spacecraft,zhou2025single} and low-thrust reachable set \cite{bando2018nonlinear,natherson2025study,lin2025spacecraft}, which has been widely applied to the analysis of space debris evolution \cite{wen2016modeling}, collision probability \cite{wen2022calculating}, missed thrust \cite{natherson2026reachabililty}, gravity assist \cite{wang2026ballistic}, pursuit-evasion game \cite{gong9103281}, and Earth-Moon transfer mission \cite{jing2026study}. To our best knowledge, the application of reachable sets to interplanetary transfer design mainly focuses on the search for planetary capture trajectories \cite{li2025expanding} and gravity assist trajectories \cite{wang2026ballistic}. Solar-sail reachable sets (especially the diffractive-sail reachable sets) for the interplanetary transfer design have not been systematically investigated. Recently, Acciarini et al. \cite{acciarini2026reachability} proposed the reflective-sail reachable sets with the maximum initial mass. However, their concept of reachable sets is constrained by the presented final target, which differs from the single-impulse reachable sets to be considered in this paper. For the single-impulse reachable sets, several scholars have proposed computational methods. Wen et al. \cite{wen2023reachable} proposed a computational method for reachable sets under the Earth-centered two-body problem with the $J_2$ perturbation using the envelope theory. Furthermore, Zhou et al. \cite{zhou2025single} proposed a computational method under arbitrary dynamics using differential algebra (DA) with adaptive domain splitting (ADS). Their methods focused on solving the boundaries of the reachable sets. Specifically, they selected the maximum impulse to obtain the reachable-set boundary. This selection is based on the linear approximation and convexity of the reachable sets, perfectly suitable for the short-time, low-impulse scenario. However, in a long-time, high-impulse scenario, such as an interplanetary transfer, the maximum impulse is insufficient to obtain the reachable-set boundary due to the pronounced nonlinear effects. Therefore, for reachable-set-based interplanetary transfer design, a wider range of impulse (departure hyperbolic excess velocity) should be considered. Based on the work of Wen et al. \cite{wen2023reachable} and Zhou et al. \cite{zhou2025single}, this paper further proposes a complementary computational method for reachable sets under arbitrary dynamics, which is suitable for the long-time, high-impulse scenario. 

The main purposes of this paper are to propose a complementary computational method for reachable sets under arbitrary dynamics and to apply it to the diffractive-sail reachable set and interplanetary transfer design. First, the diffractive-sail interplanetary transfer problem is transformed into the problem of computing single-impulse reachable sets. Then, a complementary computational method for reachable sets under arbitrary dynamics is established using DA. We select a range from 0 to the maximum departure hyperbolic excess velocity to compute the reachable sets. To ensure the accuracy and local resolution of the reachable sets, adaptive grid refinement using the quadtree method is combined with the computational method. Differing from the method proposed by Zhou et al. \cite{zhou2025single} using the ADS, the proposed quadtree framework allows different merit scores (not only the truncation error) to achieve different design targets. The proposed method is then applied to compute diffractive-sail reachable sets. The configurations of the obtained reachable sets are accordingly analyzed, providing the position information for the interplanetary transfer design. Finally, a preliminary design of interplanetary transfer (Earth-Mars transfer) is performed based on the obtained reachable sets. Specifically, the diffractive-sail reachable sets are used to generate initial guesses. The design results are analyzed and discussed, and the effects of the corresponding parameters (e.g., transfer time, diffractive angle, type of diffractive sails) on transfer characteristics are analyzed, providing further insight into parameter selection in the interplanetary transfer design. This work established a direct link between interplanetary transfer, the reachable set, and diffractive-sail dynamics.

The rest of this paper is organized as follows. Section \ref{sec2} presents the problem statement of the diffractive-sail single-impulse reachable set. Section \ref{sec3} proposes the complementary computational method for reachable sets under arbitrary dynamics and applies it to the diffractive-sail reachable set. Section \ref{sec4} performs the preliminary design of the Earth-Mars Transfers and analyzes the effects of the corresponding parameters on the transfer characteristics. Finally, conclusions are drawn in Section \ref{sec5}.

\section{Diffractive-Sail Single-Impulse Reachable Set}\label{sec2}
This section presents the foundation of this paper, including the dynamical model of a diffractive sail, the interplanetary transfer design problem, and the corresponding transformed problem of computing single-impulse reachable sets.
\subsection{Heliocentric Two-Body Problem with a Diffractive Sail}\label{subsec2.1}
For an interplanetary transfer problem using a diffractive sail, we adopt the heliocentric two-body problem with a diffractive sail as the dynamical model. The heliocentric inertial frame is used, and a set of dimensionless units is selected as follows: the length unit (LU) is set as the Sun-Earth distance (i.e., one astronomical unit), the mass unit (MU) is set as the mass of the Sun, and the time unit (TU) is set as $\sqrt {{{\text{LU}}^3}/G\left({\text{MU}}\right)}$, where $G$ denotes the gravitational constant. The values of these dimensionless units are selected from \url{https://ssd.jpl.nasa.gov/astro_par.html}. Neglecting the inclinations of the common planets around the Sun, we consider the planar problem in this paper. Therefore, the dynamical equations can be written as:
\begin{equation}
\begin{gathered}
  \dot x = u \hfill \\
  \dot y = v \hfill \\
  \dot u =  - \frac{x}{{{r^3}}} + \frac{\beta }{{2{r^3}}}\left( {x\left( {1 \pm \cos \theta } \right) - y\sin \theta } \right) \hfill \\
  \dot v =  - \frac{y}{{{r^3}}} + \frac{\beta }{{2{r^3}}}\left( {y\left( {1 \pm \cos \theta } \right) + x\sin \theta } \right) \hfill \\
  r = \sqrt {{x^2} + {y^2}}  \hfill \\ 
\end{gathered}
\label{eq1}
\end{equation}
where $\bm{X}=\left[x,\text{ }y,\text{ }u,\text{ }v\right]^\text{T}$ denotes the state in the heliocentric inertial frame, $r$ denotes the distance between the diffractive sail and the Sun, $\beta$ denotes the lightness number of the diffractive sail, and $\theta \in \left[-\frac{\pi}{2},\text{ }\frac{\pi}{2}\right]$ denotes the diffractive angle which remains constant during the trajectory propagation. According to Ref. \cite{chu2024potential}, the term $1+\cos{\theta}$ corresponds to the reflection-type diffractive sail (RDS), while the term $1-\cos{\theta}$ corresponds to the transmission-type diffractive sail (TDS). This paper addresses the solution of reachable sets for both RDS and TDS. Furthermore, we consider the cases with the constant diffractive angle during the trajectory propagation. Generally, no closed-form solution exists for Eq. \eqref{eq1}. Therefore, we propagate diffractive-sail trajectories using an adaptive Fehlberg embedded seventh/eighth-order Runge-Kutta (RKF78) method with absolute and relative tolerances set to $1\times 10^{-13}$. Then, we describe the interplanetary transfer design problem and transform it into a problem of computing single-impulse reachable sets.
\subsection{Interplanetary Transfer Problem and Reachable Set}\label{subsec2.2}
Interplanetary transfer describes a scenario where a spacecraft is launched to escape from Earth, enters a heliocentric transfer trajectory, and is finally captured by a target planet, such as Mars. In this paper, we focus on the design of the heliocentric transfer trajectory. For transfer design in the heliocentric two-body problem without the solar-sail spacecraft, methods based on the two-body closed-form solution, such as the Lambert transfer method and Hohmann transfer method \cite{battin1999introduction}, are usually used. For transfer design with a diffractive-sail spacecraft, the introduction of the SRP acceleration makes closed-form solutions absent. Therefore, the interplanetary transfer design for the diffractive sail in the planar problem can be transformed into the search for feasible solutions satisfying the specific constraints (heliocentric radius of the target planet) for Eq. \eqref{eq1}. To perform this search, a numerical method should be developed. Similar to the Earth-Moon transfer design in the multi-body problem \cite{topputo2013optimal,oshima2019low}, the design procedure can be divided into two steps: generation of initial guesses and trajectory correction. To generate initial guesses, a reachability analysis problem should be addressed. Differing from work on solar-sail transfer design assuming zero departure hyperbolic excess velocities \cite{gong2015interplanetary,firuzi2021gradient,chu2024minimum}, in this paper, we consider the diffractive-sail transfers with nonzero departure hyperbolic excess velocities. Therefore, the reachability analysis of a diffractive sail can be further transformed into a problem of computing “single-impulse” reachability sets. In particular, the impulse denotes the hyperbolic excess velocity ($v_{{\infty},\text{ }d}$; the subscript “\textit{d}” denotes quantities corresponding to the departure from Earth). In this paper, we focus on the position of reachable sets, also denoted as the reachable domain \cite{wen2023reachable}. Here, we present the mathematical definition of the single-impulse reachable set for a diffractive sail. We set the initial state of Earth in the heliocentric inertial frame to $\bm{X}_\text{Earth}=\left[1,\text{ }0,\text{ }0,\text{ }1\right]^\text{T}$. Then, the initial state $\bm{X}_d$ of the diffractive sail can be expressed as:
\begin{equation}
\begin{gathered}
   x_d = 1 \hfill \\
  y_d = 0 \hfill \\
  u_d =  v_{{\infty},\text{ }d} \cos \alpha \hfill \\
  v_d =  1 + v_{{\infty},\text{ }d} \sin \alpha \hfill \\
\end{gathered}
\label{eq2}
\end{equation}
where $v_{{\infty},\text{ }d}$ denotes the departure hyperbolic excess velocity, and $\alpha$ denotes the phase angle. In this paper, we focus on the preliminary design of the Earth-Mars transfers. Therefore, we set to $v_{{\infty},\text{ }d} \in \left[0,\text{ }5\right]\text{ }\left(\text{km/s}\right)$ and $\alpha \in \left[0,\text{ }2\pi\right)$. Moreover, to illustrate the extensibility of the method, we present an example of designing transfers to other planets (Earth-Venus transfer) in the Appendix. We consider the reachable sets with the fixed propagation time. Let $\bm{X}_f$ be the state propagated from $\bm{X}_d$ with the propagation time $T$ using Eq. \eqref{eq1}, the reachable sets $\mathcal{R}\left(T\right)$ with $T$ can be expressed as:
\begin{equation}
\mathcal{R}\left(T\right)=\{\left(x_f,\text{ }y_f\right)|\bm{X}_f=\phi_{0}^{T}\left(\bm{X}_d\right),\text{ }v_{{\infty},\text{ }d} \in \left[0,\text{ }5\right]\text{ }\left(\text{km/s}\right),\text{ }\alpha \in \left[0,\text{ }2\pi\right)\}
\label{eq3}
\end{equation}
where $\phi:\text{ }\mathbb{R}^4 \to \mathbb{R}^4,\text{ }\bm{X}_d \to \bm{X}_f$ denotes the dynamical flow of Eq. \eqref{eq1}. To obtain reachable sets, the simple grid search method can be used, i.e., dividing the grid within $v_{{\infty},\text{ }d} \in \left[0,\text{ }5\right]\text{ }\left(\text{km/s}\right)$ and $\alpha \in \left[0,\text{ }2\pi\right)$, and recording the $\left(x_f,\text{ }y_f\right)$ set through the RKF78 method. This process can be time-consuming \cite{zhou2025single}. To further improve the computational efficiency of $\mathcal{R}\left(T\right)$, following the pioneering work of Zhou et al. \cite{zhou2025single}, we further develop a complementary DA-based computational method suitable for the interplanetary transfer scenario in this paper.
\section{Differential-Algebra-Based Reachable Set}\label{sec3}
In this section, we propose a complementary DA-based computational method for reachable sets based on Ref. \cite{zhou2025single}. First, we discuss the parameter setting to compute reachable sets. Then, a DA-based computational method with adaptive grid refinement is proposed, and the corresponding results are presented and discussed.
\subsection{Parameter Setting}\label{subsec3.1}
\begin{figure}[H]
\centerline{\includegraphics[width=0.8\textwidth]{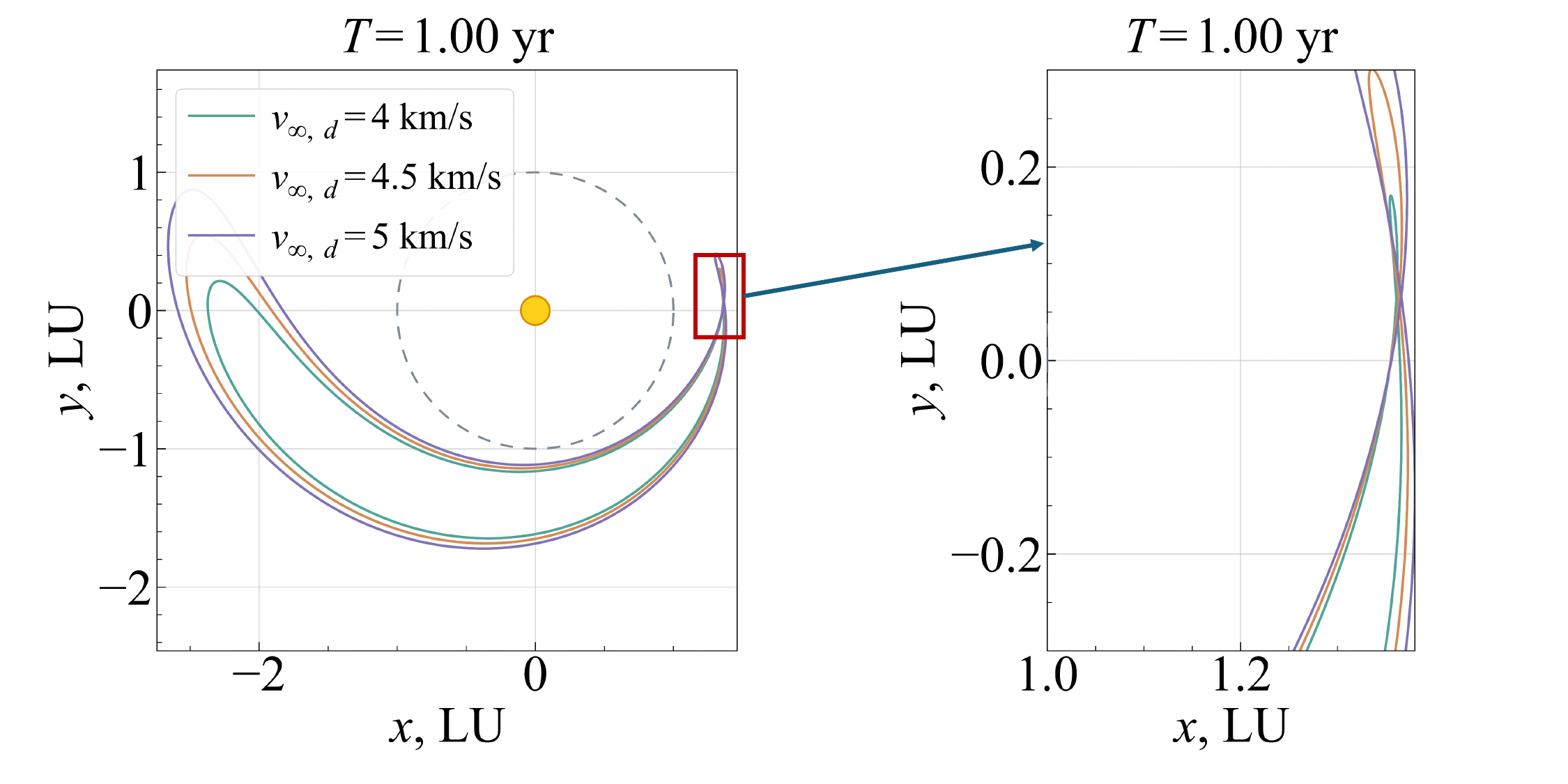}}
\caption{Equal-$v_{\infty,\text{ }d}$ contours with the fixed values of $v_{{\infty},\text{ }d}$ ($v_{{\infty},\text{ }d}=4,\text{ }4.5,\text{ }5\text{ km/s}$).}\label{fig_equal_vinf}
\end{figure}

The reachable sets considered in this paper are two-parameter reachable sets, i.e., they are determined by two parameters ($\alpha$ and $v_{{\infty},\text{ }d}$) used to compute the initial states of a diffractive sail. For single-impulse reachable sets, the maximum value of impulse ($v_{{\infty},\text{ }d}$ in this paper) usually determines their boundary \cite{wen2016modeling,wen2023reachable,zhou2025single}. This conclusion has been perfectly suitable for the reachable sets with relatively low impulse and short propagation time, in which cases the convexity (linear approximation \cite{lin2025spacecraft}) of the reachable sets is well preserved. However, for interplanetary transfers, relatively high $v_{{\infty},\text{ }d}$ (impulse) and long propagation time are required, in which cases pronounced nonlinear effects can yield a distortion of the configurations of the reachable sets. As shown in Fig. \ref{fig_equal_vinf}, setting the propagation time to 1 year, the equal-$v_{\infty,\text{ }d}$ contour (using an RDS with $\theta=\pi/3$) with $v_{{\infty},\text{ }d}=5\text{ km/s}$ do not involve the whole region with $v_{{\infty},\text{ }d}\leq 5\text{ km/s}$, as part of the equal-$v_{\infty,\text{ }d}$ contours with $v_{{\infty},\text{ }d}=4\text{ km/s}$ and $v_{{\infty},\text{ }d}=4.5\text{ km/s}$ lies outside the region inside that with $v_{{\infty},\text{ }d}=5\text{ km/s}$. This special property indicates that when computing diffractive-sail reachable sets, the grid search with respect to $v_{{\infty},\text{ }d}$ is also required (not only selecting the maximum $v_{{\infty},\text{ }d}$). Subsequently, a computational method for single-impulse reachable sets is proposed using DA and adaptive grid refinement with respect to $\alpha$ and $v_{{\infty},\text{ }d}$.
\subsection{Computational Method}\label{subsec3.2}
The computational method proposed in this paper can be divided into two aspects: DA-based propagation and adaptive grid refinement. The method is then described in detail.
\subsubsection{DA-Based Propagation}\label{subsubsec3.2.1}
Based on the discussion in Section \ref{subsec3.1}, we compute the reachable sets with adaptive grid refinement with respect to $\alpha$ and $v_{{\infty},\text{ }d}$. The ranges of $\alpha$ and $v_{{\infty},\text{ }d}$ are set to $\alpha \in \left[0,\text{ }2\pi\right)$ and $v_{{\infty},\text{ }d} \in \left[0,\text{ }5\right]\text{ }\left(\text{km/s}\right)$. With these ranges, we obtain several $\left(\alpha,\text{ }v_{{\infty},\text{ }d}\right)$ cells (detailed in Section \ref{subsubsec3.2.2}). The DA uses the Taylor polynomial to approximate the states near a reference trajectory. For a $\left(\alpha,\text{ }v_{{\infty},\text{ }d}\right)$ cell, we select the cell center, propagate the corresponding trajectory using the RKF78 method, and treat it as the reference trajectory of the corresponding cell. Then, The DA-based propagation is performed with respect to each reference trajectory, using DACEyPy 1.3.0 \footnote{\url{https://pypi.org/project/daceypy/}} in a Python 3.12.12 environment and the RKF78 method with absolute and relative tolerances set to $1\times 10^{-13}$. The final states $\bm{X}_{f,\text{ }j}$ in the corresponding cell can be expressed as:
\begin{equation}
\begin{gathered}
{x_{f,{\text{ }}j}} \approx {\mathcal{T}_{x,\text{ }j}}\left( {{\xi _{\alpha_j}},{\text{ }}{\xi _{v_{\infty ,{\text{ }d,\text{ }j}}}}} \right) = \sum\limits_{m + n \le k} {{b_{x,{\text{ }}m,{\text{ }}n,\text{ }j}}\xi_{\alpha_j} ^m\xi_{{v_{\infty ,{\text{ }}d,{\text{ }}j}}}^n} \hfill \\
{y_{f,{\text{ }}j}} \approx {\mathcal{T}_{y,\text{ }j}}\left( {{\xi _{\alpha_j}},{\text{ }}{\xi _{v_{\infty ,{\text{ }d,\text{ }j}}}}} \right) = \sum\limits_{m + n \le k} {{b_{y,{\text{ }}m,{\text{ }}n,\text{ }j}}\xi_{\alpha_j} ^m\xi_{{v_{\infty ,{\text{ }}d,{\text{ }}j}}}^n} \hfill \\
\end{gathered}
\label{eq4}
\end{equation}
\begin{equation}
\begin{gathered}
   {b_{x,{\text{ }}m,{\text{ }}n,{\text{ }}j}} = \frac{1}{{m!n!}}\left. \frac{{{\partial ^{m + n}}{x_{f,{\text{ }}j}}}}{{{\partial ^m}{\xi _{{\alpha _j}}}{\partial ^n}{\xi _{{v_{\infty ,{\text{ }}d,{\text{ }}j}}}}}}\right|_{\left(0,\text{ }0\right)}  \hfill \\
   {b_{y,{\text{ }}m,{\text{ }}n,{\text{ }}j}} = \frac{1}{{m!n!}}\left. \frac{{{\partial ^{m + n}}{y_{f,{\text{ }}j}}}}{{{\partial ^m}{\xi _{{\alpha _j}}}{\partial ^n}{\xi _{{v_{\infty ,{\text{ }}d,{\text{ }}j}}}}}}\right|_{\left(0,\text{ }0\right)} \hfill \\
\end{gathered}
\label{eq4_1}
\end{equation}
where $\mathcal{T}$ denotes the Taylor polynomial, the subscript “\textit{j}” denotes the cell label, $b$ denotes the coefficients of the Taylor polynomial, and $k$ denotes the selected highest order of the Taylor polynomial. In this paper, $k=8$. For each cell, the DA variables ${\xi _{{\alpha_j }}} \in \left[-1,\text{ }1\right]$ and $\xi_{v_{\infty ,{\text{ }d,\text{ }j}}} \in \left[-1,\text{ }1\right]$ are the normalized variables corresponding to $\alpha$ and $v_{{\infty},\text{ }d}$:
\begin{equation}
\begin{gathered}
   {\alpha_j} = {\alpha_{c,{\text{ }}j}} + \frac{1}{2}{\xi _{{\alpha_{j}}}}\left( {\max \left( {\alpha_j} \right) - \min \left( {\alpha_j} \right)} \right)  \hfill \\ 
   {\alpha_{c,{\text{ }}j}} = \frac{1}{2}\left( {\max \left( {\alpha_j} \right) + \min \left( {\alpha_j} \right)} \right)  \hfill \\
\end{gathered}
\label{eq6}
\end{equation}
\begin{equation}
\begin{gathered}
   {v_{\infty ,{\text{ }}d,{\text{ }}j}} = {v_{\infty ,{\text{ }}d,{\text{ }}c,{\text{ }}j}} + \frac{1}{2}{\xi _{{v_{\infty ,{\text{ }}d,{\text{ }}j}}}}\left( {\max \left( {{v_{\infty ,{\text{ }}d,{\text{ }}j}}} \right) - \min \left( {{v_{\infty ,{\text{ }}d,{\text{ }}j}}} \right)} \right)  \hfill \\ 
   {v_{\infty ,{\text{ }}d,{\text{ }}c,{\text{ }}j}} = \frac{1}{2}\left( {\max \left( {{v_{\infty ,{\text{ }}d,{\text{ }}j}}} \right) + \min \left( {{v_{\infty ,{\text{ }}d,{\text{ }}j}}} \right)} \right)  \hfill \\
\end{gathered}
\label{eq5}
\end{equation}
where the subscript “\textit{c}” denotes the cell center corresponding to the reference trajectory (for each reference trajectory, $\xi _{{\alpha_j }}=0$ and $\xi_{v_{\infty ,{\text{ }d,\text{ }j}}}=0$). In this paper, we used the approximated $\left(x_f,\text{ }y_f\right)$ in Eq. \eqref{eq4} to compute reachable sets. To ensure the accuracy and local resolution of reachable sets, adaptive grid refinement (the process to obtain the final $\left(\alpha,\text{ }v_{{\infty},\text{ }d}\right)$ cells) based on different merit scores is performed.
\subsubsection{Adaptive Grid Refinement}\label{subsubsec3.2.2}
In this paper, adaptive grid refinement is developed to ensure the accuracy and local resolution of reachable sets. The grid refinement is achieved by the quadtree method. We first select a $64\times 64$ grid of $\left(\alpha,\text{ }v_{{\infty},\text{ }d}\right)$, and the corresponding cells are then refined through several iterations (depths) and according to some specific criteria. The maximum depth of the quadtree method is set to 3. For each grid, we select the cell center and perform the DA propagation. Merit scores for these cells are evaluated based on the results of the DA propagation, and the cell should be refined if it satisfies the following three conditions:
\begin{enumerate}
\item The depth of the cell does not reach the maximum depth, which is set to 3;
\item The cell widths along both the $\alpha$ and $v_{{\infty},\text{ }d}$ directions exceed the tolerances, which are set to $\Delta \alpha_{\min}=\frac{2\pi}{64\times 2^3}$ and $\Delta v_{{\infty},\text{ }d\min}=\frac{5/\text{VU}}{64\times 2^3}$;
\item The corresponding merit score of this grid is not smaller than the percentile across all grids. In this paper, the percentile is set to 80.
\end{enumerate}
When the cell should be refined, the cell is bisected along both the $\alpha$ and $v_{{\infty},\text{ }d}$ directions, yielding four child cells. To ensure the accuracy and local resolution of the reachable sets, several merit scores are proposed. The merit scores adopted in this paper include two types: one type corresponding to position dispersion with respect to the reference trajectory (cell center) and local resolution, and the other type corresponding to the truncation error of the DA propagation and accuracy. The merit score corresponding to position dispersion with respect to the reference trajectory is first presented. We select the first-order terms of Taylor polynomials shown in Eq. \eqref{eq4} and estimate the boundary of the position dispersion with respect to the reference trajectory ($x_{f,\text{ }c,\text{ }j}=\mathcal{T}_{x,\text{ }j}\left(0,\text{ }0\right)$ and $y_{f,\text{ }c,\text{ }j}=\mathcal{T}_{y,\text{ }j}\left(0,\text{ }0\right)$):
\begin{equation}
\begin{gathered}
  {\hat{D}_{x,{\text{ }}j}} = {b_{x,{\text{ }}1,{\text{ }}0,\text{ }j}}{\xi _{{\alpha _j}}} + {b_{x,{\text{ 0}},{\text{ 1},\text{ }j}}}{\xi _{{v_{\infty ,{\text{ }}d,{\text{ }}j}}}}  \hfill \\
   {\hat{D}_{y,{\text{ }}j}} = {b_{y,{\text{ }}1,{\text{ }}0,\text{ }j}}{\xi _{{\alpha _j}}} + {b_{y,{\text{ 0}},{\text{ 1},\text{ }j}}}{\xi _{{v_{\infty ,{\text{ }}d,{\text{ }}j}}}}  \hfill \\
   {\xi _{{\alpha _j}}} \in \left[ { - 1,{\text{ }}1} \right]  \hfill \\ 
   {\xi _{{v_{\infty ,{\text{ }}d,{\text{ }}j}}}} \in \left[ { - 1,{\text{ }}1} \right]\hfill \\
\end{gathered}
\label{eq7}
\end{equation}
\begin{equation}
\begin{gathered}
  {\left| {{\hat{D}_{x,{\text{ }}j}}} \right|_{\max }} \leqslant \left| {{b_{x,{\text{ }}1,{\text{ }}0,\text{ }j}}} \right| + \left| {{b_{x,{\text{ 0}},{\text{ 1},\text{ }j}}}} \right|  \hfill \\
   {\left| {{\hat{D}_{y,{\text{ }}j}}} \right|_{\max }} \leqslant \left| {{b_{y,{\text{ }}1,{\text{ }}0,\text{ }j}}} \right| + \left| {{b_{y,{\text{ 0}},{\text{ 1},\text{ }j}}}} \right| \hfill \\
\end{gathered}
\label{eq8}
\end{equation}
where the superscript $\hat{\left(\cdot\right)}$ denotes the estimation value. Therefore, the first type of the merit score can be expressed as:
\begin{equation}
{S_1} = \sqrt {{{\left( {\left| {{b_{x,{\text{ }}1,{\text{ }}0,\text{ }j}}} \right| + \left| {{b_{x,{\text{ 0}},{\text{ 1},\text{ }j}}}} \right|} \right)}^2} + {{\left( {\left| {{b_{y,{\text{ }}1,{\text{ }}0,\text{ }j}}} \right| + \left| {{b_{y,{\text{ 0}},{\text{ 1},\text{ }j}}}} \right|} \right)}^2}} 
\label{eq9}
\end{equation}
A higher $S_1$ approximately indicates that the maximum final position dispersion with respect to the reference trajectory of this cell is relatively large, which indicates the relatively bad local resolution. The cell should be refined to reduce the final position dispersion with respect to the reference trajectory. Then, the merit score corresponding to the truncation error of the DA propagation is presented. Notably, this merit score borrows the idea of the ADS \cite{wittig2015ads,zhou2025single}. The sum of the absolute values of the coefficients of the Taylor polynomials for each order $l$ of each cell can be presented as follows:
\begin{equation}
\begin{gathered}
H_{x,{\text{ }}j}^{\left( l \right)} = \sum\limits_{m + n = l} {\left| {{b_{x,{\text{ }}m,{\text{ }}n,\text{ }j}}} \right|}   \hfill \\ 
   H_{y,{\text{ }}j}^{\left( l \right)} = \sum\limits_{m + n = l} {\left| {{b_{y,{\text{ }}m,{\text{ }}n,\text{ }j}}} \right|}  \hfill \\
\end{gathered}
\label{eq10}
\end{equation}
Assuming that the absolute values of the coefficients of the Taylor polynomials have an exponential decay when the order $l$ increases, we have the following approximation relationship:
\begin{equation}
\begin{gathered}
 \ln H_{x,{\text{ }}j}^{\left( l \right)} \approx {q_{x,{\text{ }}j}} + {p_{x,{\text{ }}j}}l  \hfill \\
  \ln H_{y,{\text{ }}j}^{\left( l \right)} \approx {q_{y,{\text{ }}j}} + {p_{y,{\text{ }}j}}l \hfill \\
\end{gathered}
\label{eq11}
\end{equation}
We select the last four orders of the Taylor polynomials shown in Eq. \eqref{eq4}, and adopt the least-squares method to fit $q_{x,\text{ }j}$, $p_{x,\text{ }j}$, $q_{y,\text{ }j}$, and $p_{x,\text{ }j}$. Then, the sum of the $\left(k+1\right)$-order coefficients can be estimated as:
\begin{equation}
\begin{gathered}
 \hat H_{x,{\text{ }}j}^{\left( {k + 1} \right)} = \exp \left[ {{q_{x,j}} + {p_{x,j}}\left( {k + 1} \right)} \right]   \hfill \\
  \hat H_{y,{\text{ }}j}^{\left( {k + 1} \right)} = \exp \left[ {{q_{y,j}} + {p_{y,j}}\left( {k + 1} \right)} \right]  \hfill \\
\end{gathered}
\label{eq12}
\end{equation}
Then, the boundary of the truncation error can be estimated as (${\xi _{{\alpha_j }}} \in \left[-1,\text{ }1\right]$ and $\xi_{v_{\infty,\text{ }j,{\text{ }d}}} \in \left[-1,\text{ }1\right]$):
\begin{equation}
\begin{gathered}
   \left| {{{\hat {Er}}_{x,{\text{ }}j}}} \right| \approx \hat H_{x,{\text{ }}j}^{\left( {k + 1} \right)}  \hfill \\
  \left| {{{\hat {Er}}_{y,{\text{ }}j}}} \right| \approx \hat H_{y,{\text{ }}j}^{\left( {k + 1} \right)} \hfill \\
\end{gathered}
\label{eq13}
\end{equation}
Then, the second type of the merit score is selected as:
\begin{equation}
{S_2} = \sqrt {{{\left( \hat H_{x,{\text{ }}j}^{\left( {k + 1} \right)} \right)}^2} + {{\left( \hat H_{y,{\text{ }}j}^{\left( {k + 1} \right)} \right)}^2}} 
\label{eq14}
\end{equation}
A higher $S_2$ approximately indicates a higher maximum truncation error of the DA propagation for this cell. The cell should be refined to decrease the maximum truncation error of the DA propagation. By combining the adaptive grid refinement and the DA propagation, the reachable sets can be further computed. The computation procedure can be divided into three steps as follows:
\begin{enumerate}
\item Using the quadtree method to perform adaptive grid refinement and obtain the corresponding cells;
\item Selecting 25 samples (including endpoints) along both the $\alpha$ and $v_{{\infty},\text{ }d}$ directions for each cell (i.e., $25\times 25$ samples for each cell) and removing the duplicate samples;
\item Using the cell center to perform the DA propagation and evaluating $x_{f,\text{ }j}$ and $y_{f,\text{ }j}$ of samples through the Taylor polynomials shown in Eq. \eqref{eq4}.
\end{enumerate}
In this paper, we set the propagation time $T$ to 0.25, 0.50, 0.75, 1.00, 1.25, 1.50, 1.75, and 2.00 yr to compute the reachable sets. The diffractive angle $\theta$ is set to $\pi/6$, $\pi/3$, and $\pi/2$, and two types of diffractive sails (RDS and TDS) are investigated. Through the aforementioned three steps, we obtain the reachable sets using two types of merit scores. To further evaluate the design performance of these two types of merit scores, a global evaluation is performed.
\subsubsection{Global Evaluation}\label{subsubsec3.2.3}
We prepare the samples for the global evaluation. The prepared samples combine a $128\times128$ grid of $\left(\alpha,\text{ }v_{{\infty},\text{ }d}\right)$, 4096 Sobol points, and 250 uniformly spaced samples along both the $\alpha$ and $v_{{\infty},\text{ }d}$ directions. The Sobol sequence is generated using a fixed seed of 2024. The duplicate samples are further removed. To quantitatively evaluate the performance of the proposed method using two types of merit scores, two corresponding merits are proposed, including the final position dispersion with respect to the reference trajectory and the error of DA propagation with respect to the RKF78 integration. The final position dispersion with respect to the reference trajectory can be computed as:
\begin{equation}
D_f = \sqrt {{{\left( {{x_{f,{\text{ }}sample,{\text{ DA}}}} - {x_{f,{\text{ }}c,{\text{ DA}}}}} \right)}^2} + {{\left( {{y_{f,{\text{ }}sample,{\text{ DA}}}} - {y_{f,{\text{ }}c,{\text{ DA}}}}} \right)}^2}} 
\label{eq15}
\end{equation}
where the subscript “\textit{sample}” denotes the quantities corresponding to the selected samples for the global evaluation, while “\textit{c}” denotes the quantities corresponding to the reference trajectory (propagated from the cell center), and “DA” denotes the quantities obtained from the DA propagation. A lower maximum $D_f$ means a finer worst-case local resolution of the reachable set. Consequently, the intersection points between the reachable sets and the constraint region (e.g., the Mars orbit in this paper) are less likely to be overlooked due to locally sparse sampling in Section \ref{subsubsec3.2.2}. The error of DA propagation with respect to the RKF78 integration can be computed as:
\begin{equation}
E{r_f} = \sqrt {{{\left( {{x_{f,{\text{ }}sample,{\text{ DA}}}} - {x_{f,{\text{ }}sample,{\text{ RKF78}}}}} \right)}^2} + {{\left( {{y_{f,{\text{ }}sample,{\text{ DA}}}} - {y_{f,{\text{ }}sample,{\text{ RKF78}}}}} \right)}^2}} 
\label{eq16}
\end{equation}
where the subscript “RKF78” denotes the quantities obtained from the RKF78 integration. A lower value of the maximum ${Er}_f$ means a higher accuracy of the reachable-set approximation using DA. Linking with Section \ref{subsubsec3.2.2}, the method using $S_1$ is targeted to achieve a lower maximum $D_f$; while the method using $S_2$ is targeted to achieve a lower maximum ${Er}_f$. Then, the computational results using the proposed method are presented and analyzed.  

\textit{Remark 1}: Compared with the simple grid search method, the adaptive grid refinement utilizes some dynamical information (e.g., the estimated dispersion with respect to the reference trajectory that indicates the local stretching of the final position under the diffractive-sail dynamics), which results in a finer worst-case local resolution of the reachable set. Furthermore, DA provides a surrogate-model-based computation of reachable sets, improving computational efficiency compared with direct numerical integration under the same conditions.

\textit{Remark 2}: Compared with the method proposed by Zhou et al. \cite{zhou2025single}, two key differences can be distinguished: on one hand, instead of computing the reachable-set boundary by selecting the maximum impulse, this paper scan the whole range of $v_{\infty,\text{ }d}$, which is more suitable for the high-impulse, long-time transfers considered in this paper; on the other hand, differing from the ADS, the proposed quadtree framework allows different merit scores (not only the truncation error) to achieve different design targets, i.e., to achieve a finer worst-case local resolution of the reachable set, $S_1$ can be adopted, while $S_2$ can be adopted to achieve a finer accuracy of the reachable-set approximation. The proposed method can also be extended to other merit scores.

\textit{Remark 3}: Similar to the method proposed by Zhou et al. \cite{zhou2025single}, the proposed computational method is also suitable for the computation of reachable sets under other dynamical models because DA and the adaptive grid refinement are not limited by the diffractive-sail dynamics.

\subsection{Diffractive-Sail Reachable Sets}\label{subsec3.3}
\begin{figure}[H]
\centerline{\includegraphics[width=0.6\textwidth]{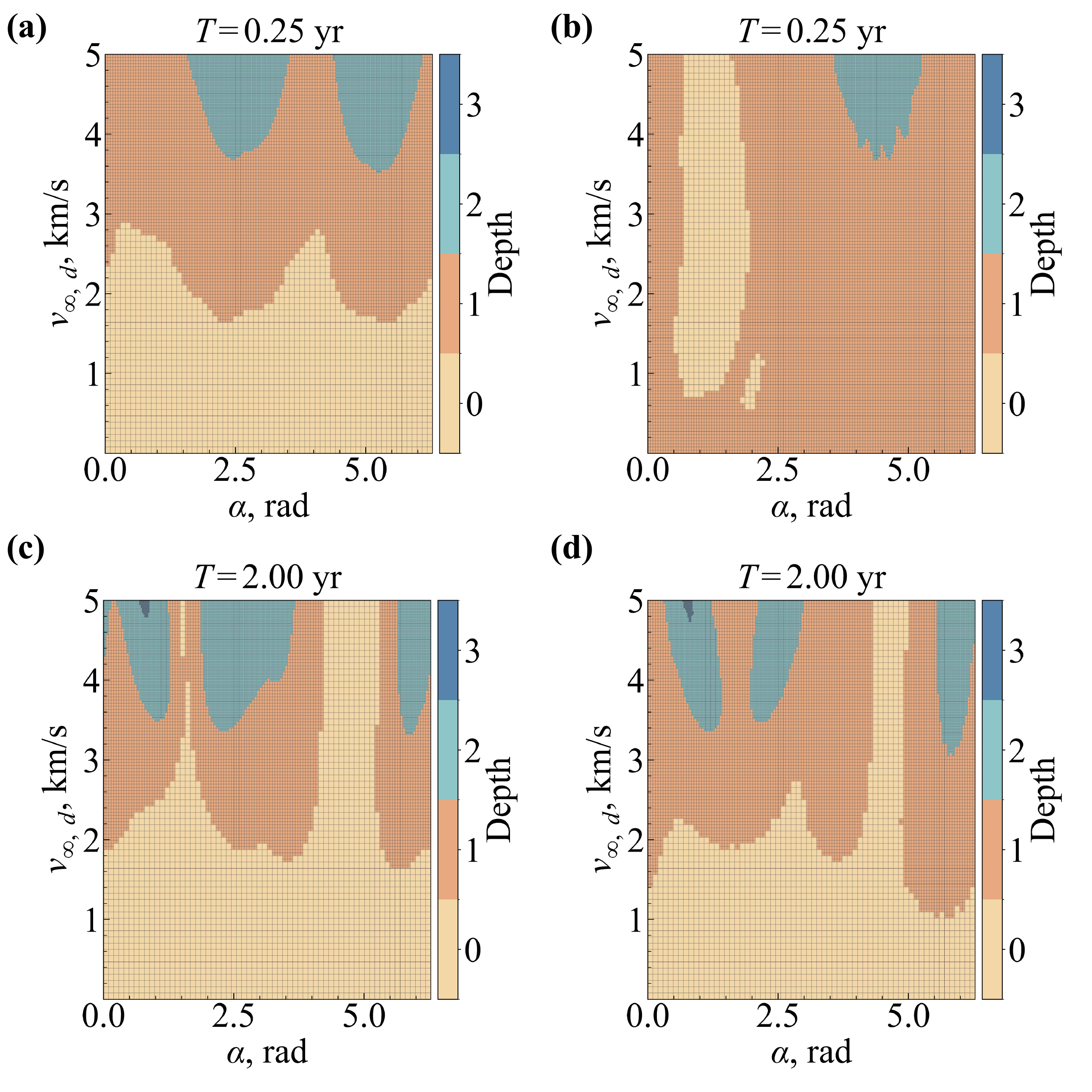}}
\caption{Examples of the adaptive grid refinement. (a) $T=0.25\text{ yr}$, $S_1$; (b) $T=0.25\text{ yr}$, $S_2$; (c) $T=2.00\text{ yr}$, $S_1$; (d) $T=2.00\text{ yr}$, $S_2$.}\label{fig_mesh}
\end{figure}

\begin{figure}[H]
\centerline{\includegraphics[width=0.6\textwidth]{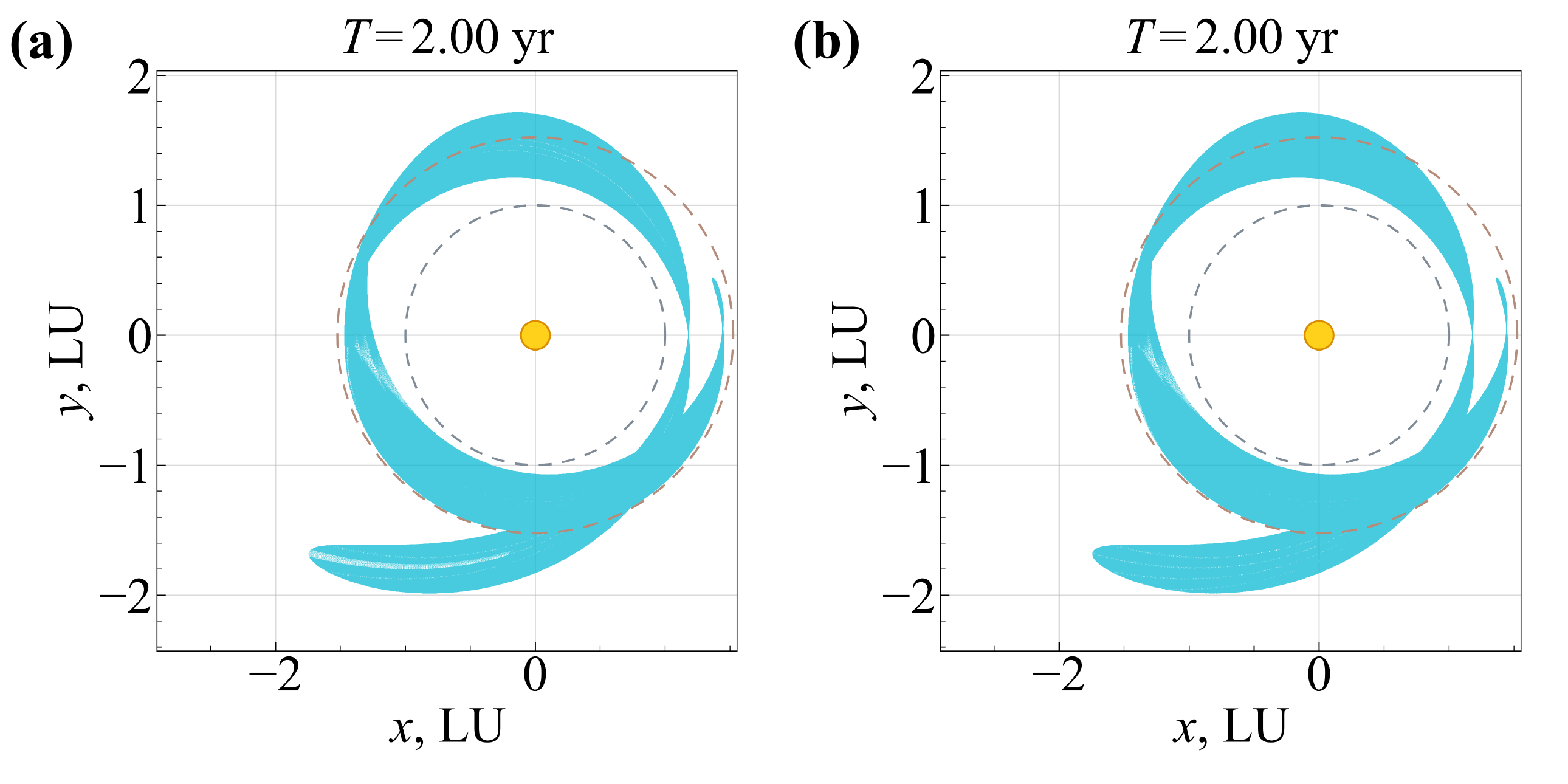}}
\caption{The differences between reachable sets obtained using $S_1$ and $S_2$.}\label{fig_S1_S2}
\end{figure}

Part of the results of the adaptive grid refinement (using an RDS with $\theta=\pi/6$) is presented in Fig. \ref{fig_mesh}. For the adaptive grid refinement, a larger depth implies that the corresponding cell repeatedly obtains relatively high merit scores during the refinement process. From Fig. \ref{fig_mesh}, it can be found that different types of merit scores ($S_1/S_2$) determine different results, as $S_1$ trends to refine the cells with relatively high $v_{\infty,\text{ }d}$ (cells with a larger depth can be observed in the region with relatively high $v_{\infty,\text{ }d}$). Meanwhile, $T$ also affects the results of the adaptive grid refinement, indicating different dynamical properties within different values of $T$. A systematic comparison of method using $S_1$ and $S_2$ in terms of the maximum $D_f$ and ${Er}_f$ of the global evaluation samples for all cases considered in this paper is performed and the corresponding comparison results are presented in Tables \ref{table_RDS}-\ref{table_TDS} (when $\theta=\pi/2$, the dynamical equations of RDS and TDS are equivalent). From Tables \ref{table_RDS}-\ref{table_TDS}, it can be observed that the targets of designing two types of merit scores ($S_1$ to reduce the maximum $D_f$ of the global evaluation while $S_2$ is designed to reduce the maximum ${Er}_f$ of the global evaluation) are generally achieved, as the maximum values of $D_f$ of results obtained using $S_1$ are lower than those of results obtained using $S_2$ for all cases, and the maximum values of ${Er}_f$ of results obtained using $S_2$ are lower than those of results obtained using $S_1$ for most cases. An example of the reachable sets (using an RDS with $\theta=\pi/6$) obtained from methods using $S_1$ and $S_2$ is presented in Fig. \ref{fig_S1_S2}. Samples in the reachable set obtained using $S_1$ are typically distributed densely in the near-boundary region of the reachable set, and are distributed relatively sparsely in some interior regions. However, samples in the reachable set obtained using $S_2$ are distributed densely not only in the near-boundary region of the reachable set, but also in some interior regions. Subsequently, we focus on the results of reachable sets obtained using $S_1$.

\begin{table}[!htb]
\caption{Merit for the computational methods using two types of merit scores using an RDS}\label{table_RDS}%
\centering
\renewcommand{\arraystretch}{1.5}
\begin{tabular}{@{}lllll@{}}
\hline
$\theta$ & $T,\text{ yr}$  & $S_1/S_2$ & ${Er}_{f,\text{ }\max}$ & ${D}_{f,\text{ }\max}$\\
\hline
$\pi/6$    & 0.25   & $S_1$ & $5.10\times 10^{-14}$ & $8.81\times10^{-3}$    \\
$\pi/6$   & 0.25   & $S_2$ & $5.10\times 10^{-14}$ & $1.54\times10^{-2}$    \\
$\pi/6$    & 0.50   & $S_1$ & $1.27\times 10^{-12}$ & $2.88\times10^{-2}$    \\
$\pi/6$    & 0.50   & $S_2$ & $6.91\times 10^{-13}$ & $4.00\times10^{-2}$    \\
$\pi/6$    & 0.75   & $S_1$ & $5.92\times 10^{-12}$ & $4.74\times10^{-2}$    \\
$\pi/6$    & 0.75   & $S_2$ & $3.06\times 10^{-12}$ & $6.56\times10^{-2}$    \\
$\pi/6$    & 1.00   & $S_1$ & $1.27\times 10^{-11}$ & $6.08\times10^{-2}$    \\
$\pi/6$    & 1.00   & $S_2$ & $7.00\times 10^{-12}$ & $9.35\times10^{-2}$    \\
$\pi/6$    & 1.25   & $S_1$ & $1.65\times 10^{-11}$ & $7.48\times10^{-2}$    \\
$\pi/6$    & 1.25   & $S_2$ & $9.94\times 10^{-12}$ & $1.22\times10^{-1}$    \\
$\pi/6$    & 1.50   & $S_1$ & $1.15\times 10^{-11}$ & $8.69\times10^{-2}$    \\
$\pi/6$    & 1.50   & $S_2$ & $7.45\times 10^{-12}$ & $1.24\times10^{-1}$    \\
$\pi/6$    & 1.75   & $S_1$ & $2.43\times 10^{-11}$ & $9.88\times10^{-2}$    \\
$\pi/6$    & 1.75   & $S_2$ & $1.71\times 10^{-11}$ & $1.19\times10^{-1}$    \\
$\pi/6$    & 2.00   & $S_1$ & $4.23\times 10^{-11}$ & $1.07\times10^{-1}$    \\
$\pi/6$    & 2.00   & $S_2$ & $3.54\times 10^{-11}$ & $1.41\times10^{-1}$    \\
$\pi/3$    & 0.25   & $S_1$ & $5.00\times 10^{-14}$ & $8.84\times10^{-3}$    \\
$\pi/3$   & 0.25   & $S_2$ & $5.12\times 10^{-14}$ & $1.54\times10^{-2}$    \\
$\pi/3$    & 0.50   & $S_1$ & $1.18\times 10^{-12}$ & $2.88\times10^{-2}$    \\
$\pi/3$    & 0.50   & $S_2$ & $6.16\times 10^{-13}$ & $3.91\times10^{-2}$    \\
$\pi/3$    & 0.75   & $S_1$ & $5.00\times 10^{-12}$ & $4.64\times10^{-2}$    \\
$\pi/3$    & 0.75   & $S_2$ & $2.58\times 10^{-12}$ & $6.43\times10^{-2}$    \\
$\pi/3$    & 1.00   & $S_1$ & $1.06\times 10^{-11}$ & $5.95\times10^{-2}$    \\
$\pi/3$    & 1.00   & $S_2$ & $5.85\times 10^{-12}$ & $9.04\times10^{-2}$    \\
$\pi/3$    & 1.25   & $S_1$ & $1.93\times 10^{-11}$ & $7.23\times10^{-2}$    \\
$\pi/3$    & 1.25   & $S_2$ & $8.54\times 10^{-12}$ & $1.16\times10^{-1}$    \\
$\pi/3$    & 1.50   & $S_1$ & $2.04\times 10^{-11}$ & $8.45\times10^{-2}$    \\
$\pi/3$    & 1.50   & $S_2$ & $1.06\times 10^{-11}$ & $1.42\times10^{-1}$    \\
$\pi/3$    & 1.75   & $S_1$ & $1.28\times 10^{-11}$ & $9.70\times10^{-2}$    \\
$\pi/3$    & 1.75   & $S_2$ & $7.33\times 10^{-12}$ & $1.41\times10^{-1}$    \\
$\pi/3$    & 2.00   & $S_1$ & $1.90\times 10^{-11}$ & $1.04\times10^{-1}$    \\
$\pi/3$    & 2.00   & $S_2$ & $1.47\times 10^{-11}$ & $1.27\times10^{-1}$    \\
\hline
\end{tabular}
\end{table}

\begin{table}[!htb]
\caption{Merit for the computational methods using two types of merit scores using an RDS (continued)}\label{table_RDS_1}%
\centering
\renewcommand{\arraystretch}{1.5}
\begin{tabular}{@{}lllll@{}}
\hline
$\theta$ & $T,\text{ yr}$  & $S_1/S_2$ & ${Er}_{f,\text{ }\max}$ & ${D}_{f,\text{ }\max}$\\
\hline
$\pi/2$    & 0.25   & $S_1$ & $4.99\times 10^{-14}$ & $8.91\times10^{-3}$    \\
$\pi/2$    & 0.25   & $S_2$ & $5.43\times 10^{-14}$ & $1.55\times10^{-2}$    \\
$\pi/2$    & 0.50   & $S_1$ & $1.22\times 10^{-12}$ & $2.84\times10^{-2}$    \\
$\pi/2$    & 0.50   & $S_2$ & $6.32\times 10^{-13}$ & $3.90\times10^{-2}$    \\
$\pi/2$    & 0.75   & $S_1$ & $4.97\times 10^{-12}$ & $4.61\times10^{-2}$    \\
$\pi/2$    & 0.75   & $S_2$ & $2.58\times 10^{-12}$ & $6.43\times10^{-2}$    \\
$\pi/2$    & 1.00   & $S_1$ & $1.21\times 10^{-11}$ & $5.85\times10^{-2}$    \\
$\pi/2$    & 1.00   & $S_2$ & $5.74\times 10^{-12}$ & $8.95\times10^{-2}$    \\
$\pi/2$    & 1.25   & $S_1$ & $1.90\times 10^{-11}$ & $6.98\times10^{-2}$    \\
$\pi/2$    & 1.25   & $S_2$ & $8.44\times 10^{-12}$ & $1.13\times10^{-1}$    \\
$\pi/2$    & 1.50   & $S_1$ & $1.82\times 10^{-11}$ & $8.07\times10^{-2}$    \\
$\pi/2$    & 1.50   & $S_2$ & $1.06\times 10^{-11}$ & $1.37\times10^{-1}$    \\
$\pi/2$    & 1.75   & $S_1$ & $1.29\times 10^{-11}$ & $9.14\times10^{-2}$    \\
$\pi/2$    & 1.75   & $S_2$ & $7.55\times 10^{-12}$ & $1.38\times10^{-1}$    \\
$\pi/2$    & 2.00   & $S_1$ & $1.73\times 10^{-11}$ & $1.02\times10^{-1}$    \\
$\pi/2$    & 2.00   & $S_2$ & $8.75\times 10^{-12}$ & $1.23\times10^{-1}$    \\
\hline
\end{tabular}
\end{table}

\begin{table}[!htb]
\caption{Merit for the computational methods using two types of merit scores using a TDS}\label{table_TDS}%
\centering
\renewcommand{\arraystretch}{1.5}
\begin{tabular}{@{}lllll@{}}
\hline
$\theta$ & $T,\text{ yr}$  & $S_1/S_2$ & ${Er}_{f,\text{ }\max}$ & ${D}_{f,\text{ }\max}$\\
\hline
$\pi/6$    & 0.25   & $S_1$ & $4.76\times 10^{-14}$ & $9.04\times10^{-3}$    \\
$\pi/6$    & 0.25   & $S_2$ & $5.93\times 10^{-14}$ & $1.57\times10^{-2}$    \\
$\pi/6$    & 0.50   & $S_1$ & $2.60\times 10^{-12}$ & $2.90\times10^{-2}$    \\
$\pi/6$    & 0.50   & $S_2$ & $1.63\times 10^{-12}$ & $4.07\times10^{-2}$    \\
$\pi/6$    & 0.75   & $S_1$ & $8.48\times 10^{-12}$ & $4.64\times10^{-2}$    \\
$\pi/6$    & 0.75   & $S_2$ & $4.63\times 10^{-12}$ & $6.77\times10^{-2}$    \\
$\pi/6$    & 1.00   & $S_1$ & $1.96\times 10^{-11}$ & $5.81\times10^{-2}$    \\
$\pi/6$    & 1.00   & $S_2$ & $8.64\times 10^{-12}$ & $9.34\times10^{-2}$    \\
$\pi/6$    & 1.25   & $S_1$ & $2.97\times 10^{-11}$ & $6.99\times10^{-2}$    \\
$\pi/6$    & 1.25   & $S_2$ & $1.17\times 10^{-11}$ & $9.99\times10^{-2}$    \\
$\pi/6$    & 1.50   & $S_1$ & $2.21\times 10^{-11}$ & $8.06\times10^{-2}$    \\
$\pi/6$    & 1.50   & $S_2$ & $1.09\times 10^{-11}$ & $1.24\times10^{-1}$    \\
$\pi/6$    & 1.75   & $S_1$ & $3.33\times 10^{-11}$ & $9.33\times10^{-2}$    \\
$\pi/6$    & 1.75   & $S_2$ & $3.16\times 10^{-11}$ & $1.15\times10^{-1}$    \\
$\pi/6$    & 2.00   & $S_1$ & $5.62\times 10^{-11}$ & $9.94\times10^{-2}$    \\
$\pi/6$    & 2.00   & $S_2$ & $4.80\times 10^{-11}$ & $1.28\times10^{-1}$    \\
$\pi/3$    & 0.25   & $S_1$ & $5.91\times 10^{-14}$ & $9.02\times10^{-3}$    \\
$\pi/3$    & 0.25   & $S_2$ & $5.91\times 10^{-14}$ & $1.56\times10^{-2}$    \\
$\pi/3$    & 0.50   & $S_1$ & $1.41\times 10^{-12}$ & $2.79\times10^{-2}$    \\
$\pi/3$    & 0.50   & $S_2$ & $7.11\times 10^{-13}$ & $3.95\times10^{-2}$    \\
$\pi/3$    & 0.75   & $S_1$ & $5.71\times 10^{-12}$ & $4.65\times10^{-2}$    \\
$\pi/3$    & 0.75   & $S_2$ & $2.96\times 10^{-12}$ & $6.54\times10^{-2}$    \\
$\pi/3$    & 1.00   & $S_1$ & $1.44\times 10^{-11}$ & $5.75\times10^{-2}$    \\
$\pi/3$    & 1.00   & $S_2$ & $6.57\times 10^{-12}$ & $9.05\times10^{-2}$    \\
$\pi/3$    & 1.25   & $S_1$ & $2.20\times 10^{-11}$ & $6.89\times10^{-2}$    \\
$\pi/3$    & 1.25   & $S_2$ & $9.67\times 10^{-12}$ & $1.14\times10^{-1}$    \\
$\pi/3$    & 1.50   & $S_1$ & $1.59\times 10^{-11}$ & $7.83\times10^{-2}$    \\
$\pi/3$    & 1.50   & $S_2$ & $9.13\times 10^{-12}$ & $1.17\times10^{-1}$    \\
$\pi/3$    & 1.75   & $S_1$ & $1.66\times 10^{-11}$ & $8.84\times10^{-2}$    \\
$\pi/3$    & 1.75   & $S_2$ & $7.29\times 10^{-12}$ & $1.41\times10^{-1}$    \\
$\pi/3$    & 2.00   & $S_1$ & $2.33\times 10^{-11}$ & $9.92\times10^{-2}$    \\
$\pi/3$    & 2.00   & $S_2$ & $1.75\times 10^{-11}$ & $1.21\times10^{-1}$    \\
\hline
\end{tabular}
\end{table}

\begin{figure}[H]
\centerline{\includegraphics[width=0.96\textwidth]{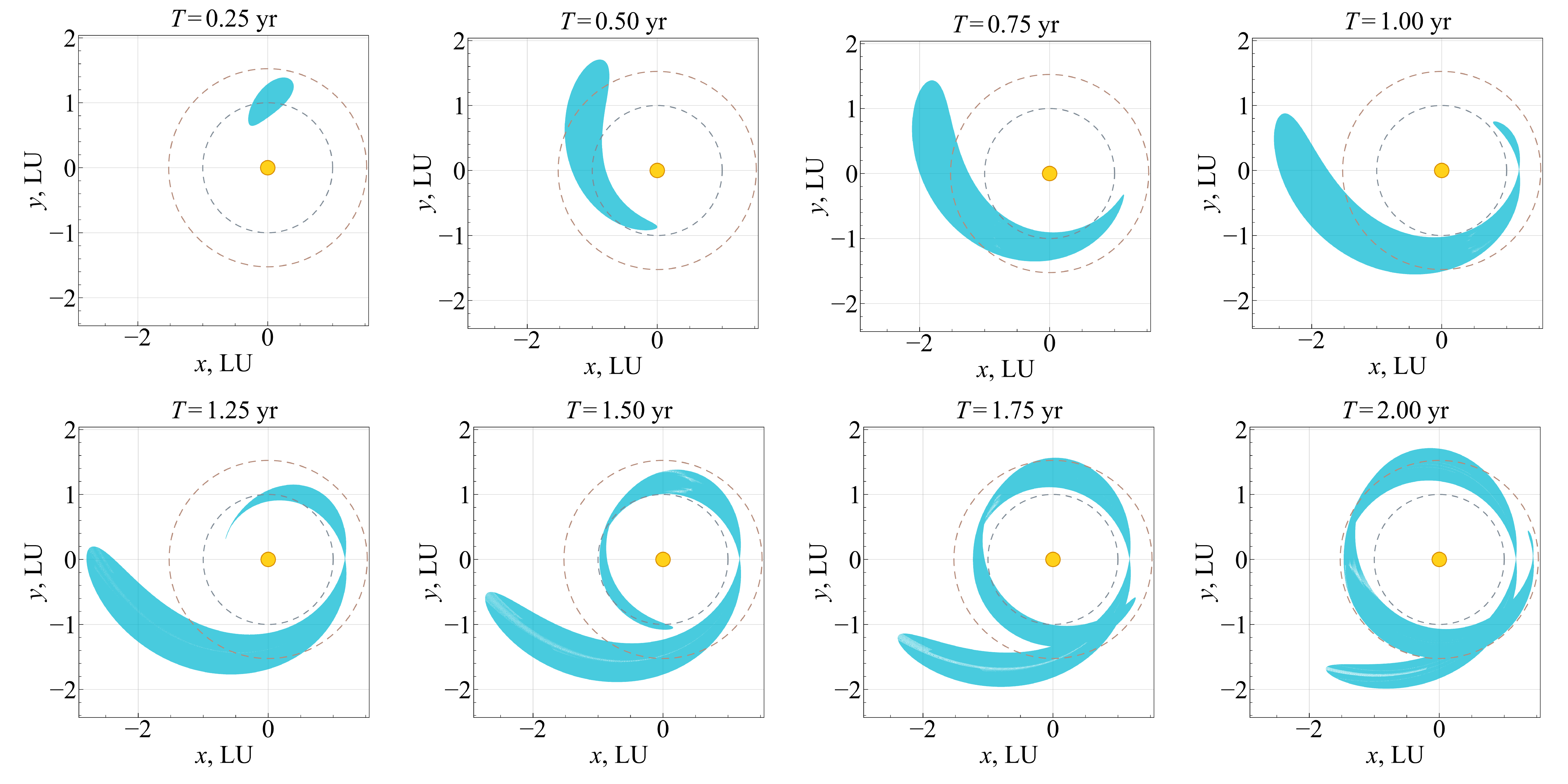}}
\caption{Configurations of reachable sets (RDS, $\theta=\pi/6$)}\label{fig_RDS_30}
\end{figure}

\begin{figure}[H]
\centerline{\includegraphics[width=0.96\textwidth]{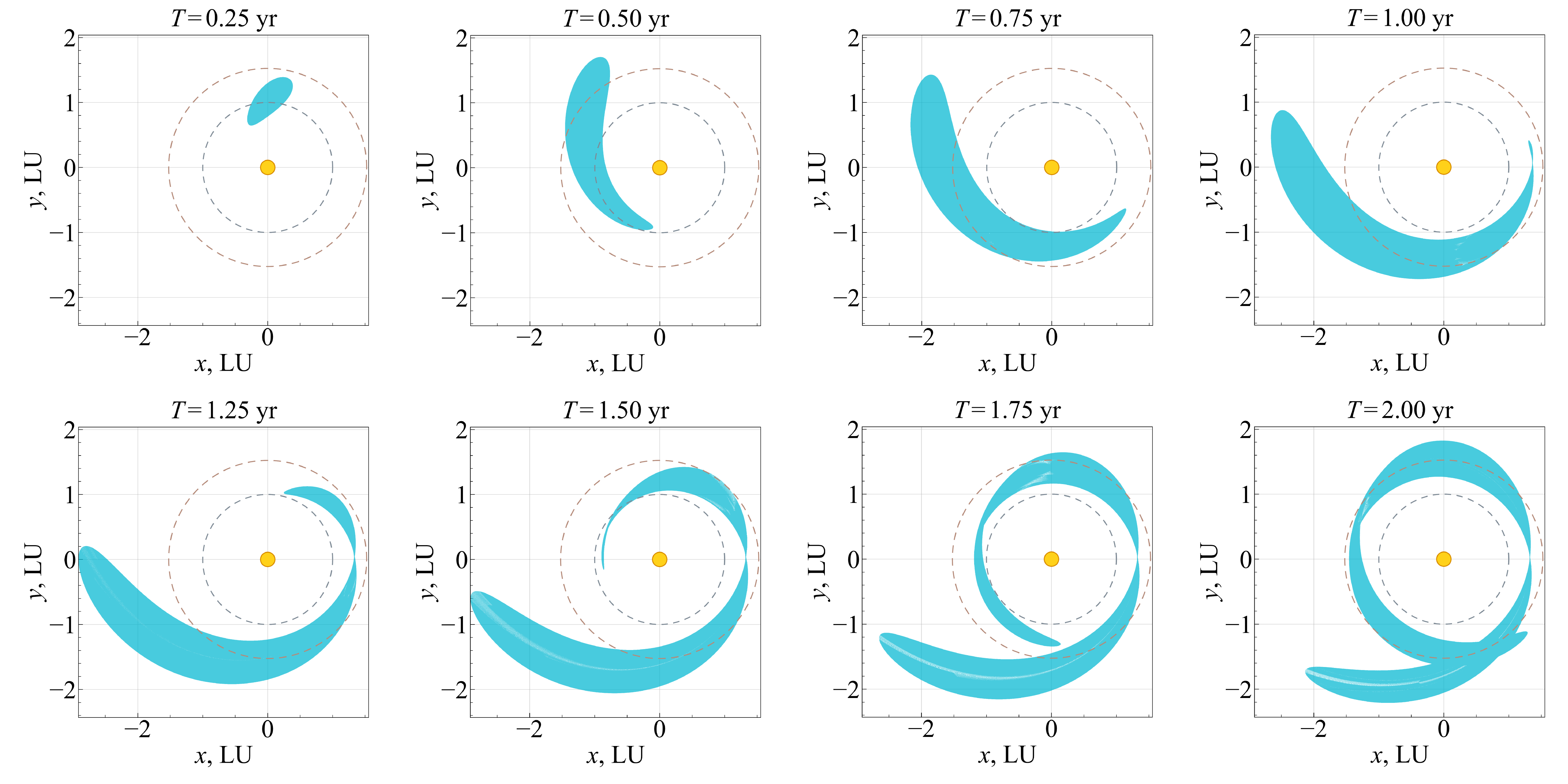}}
\caption{Configurations of reachable sets (RDS, $\theta=\pi/3$)}\label{fig_RDS_60}
\end{figure}

\begin{figure}[H]
\centerline{\includegraphics[width=0.96\textwidth]{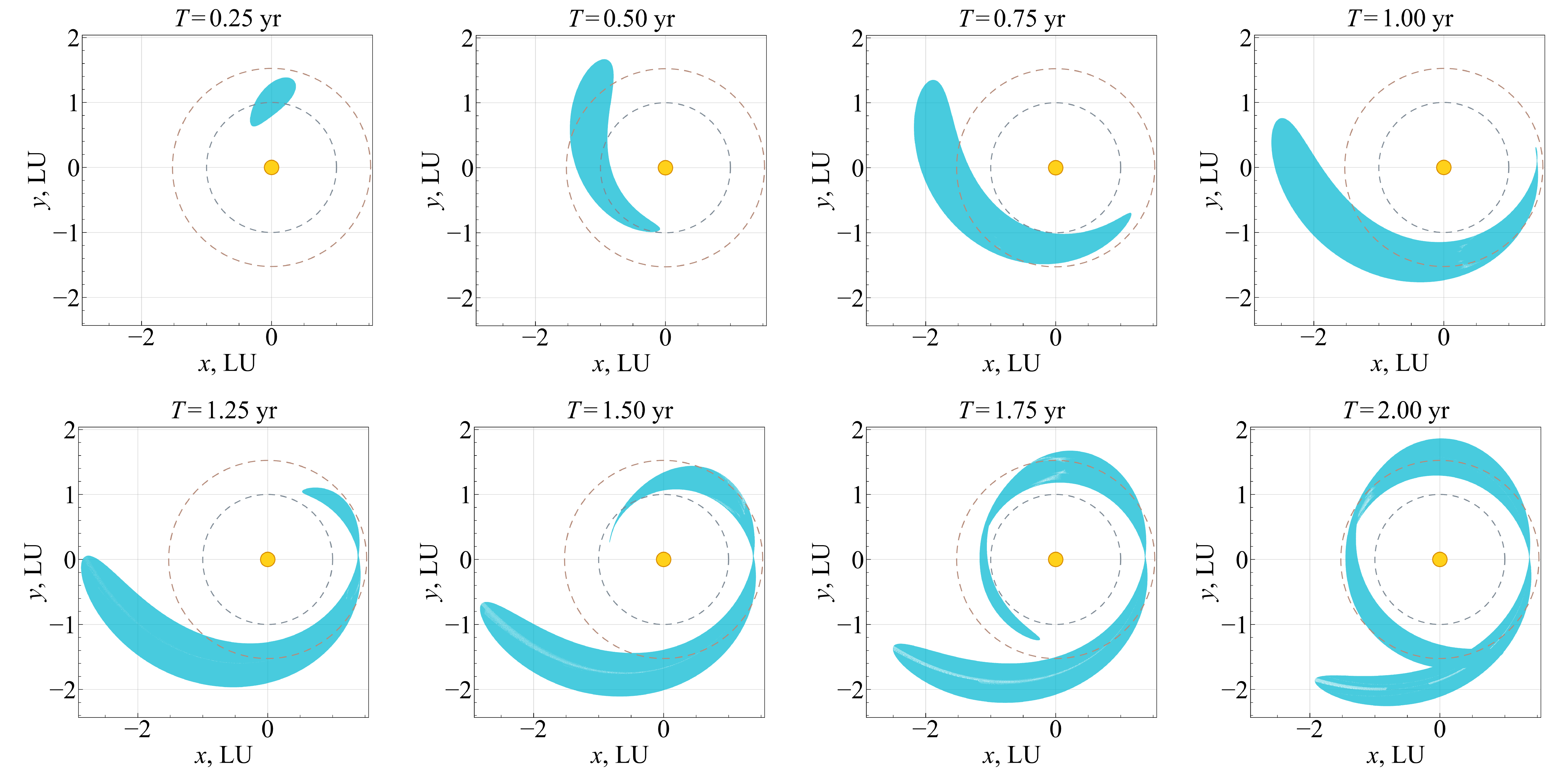}}
\caption{Configurations of reachable sets (RDS, $\theta=\pi/2$)}\label{fig_RDS_90}
\end{figure}

\begin{figure}[H]
\centerline{\includegraphics[width=0.96\textwidth]{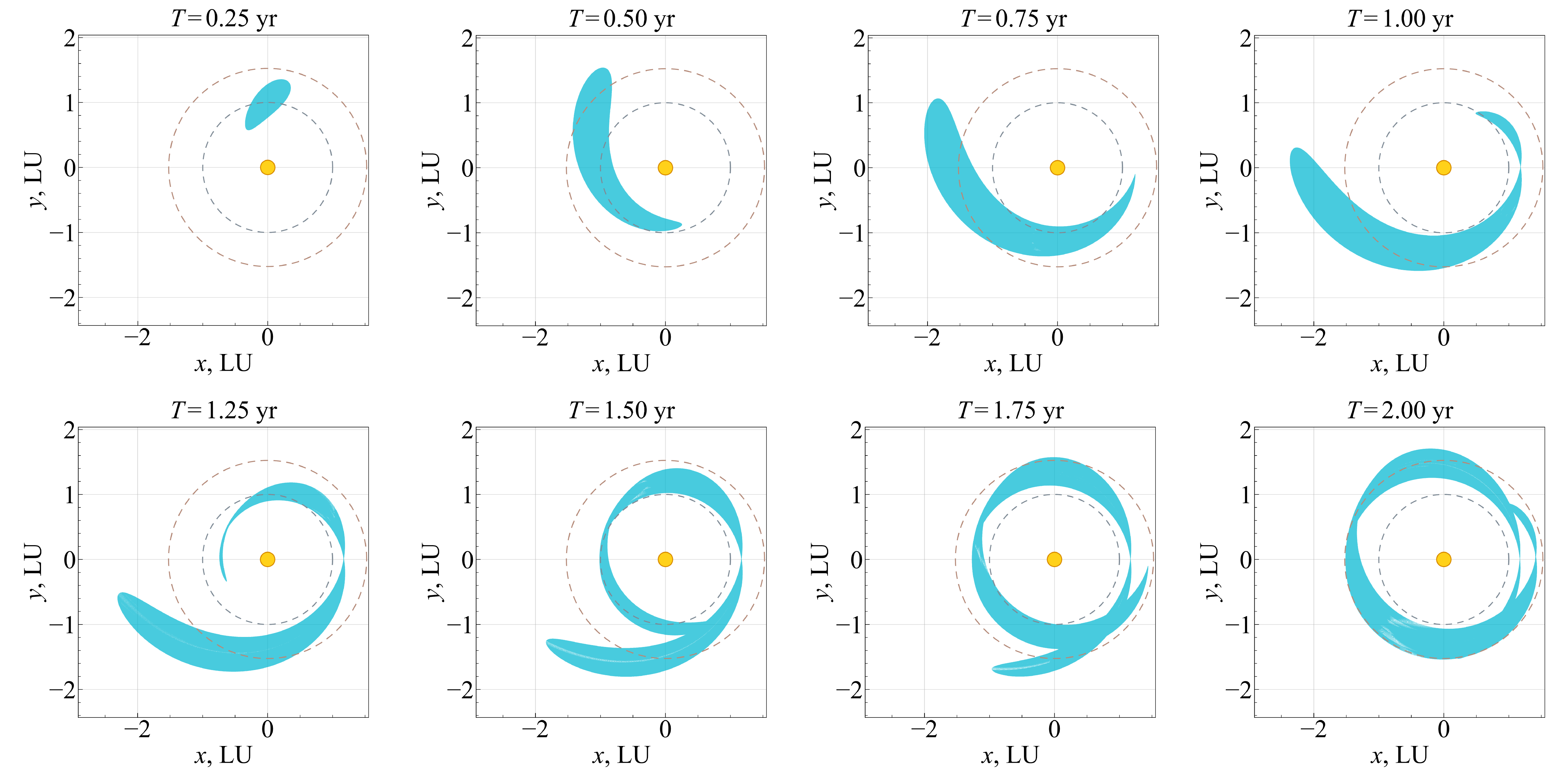}}
\caption{Configurations of reachable sets (TDS, $\theta=\pi/6$)}\label{fig_TDS_30}
\end{figure}

\begin{figure}[H]
\centerline{\includegraphics[width=0.96\textwidth]{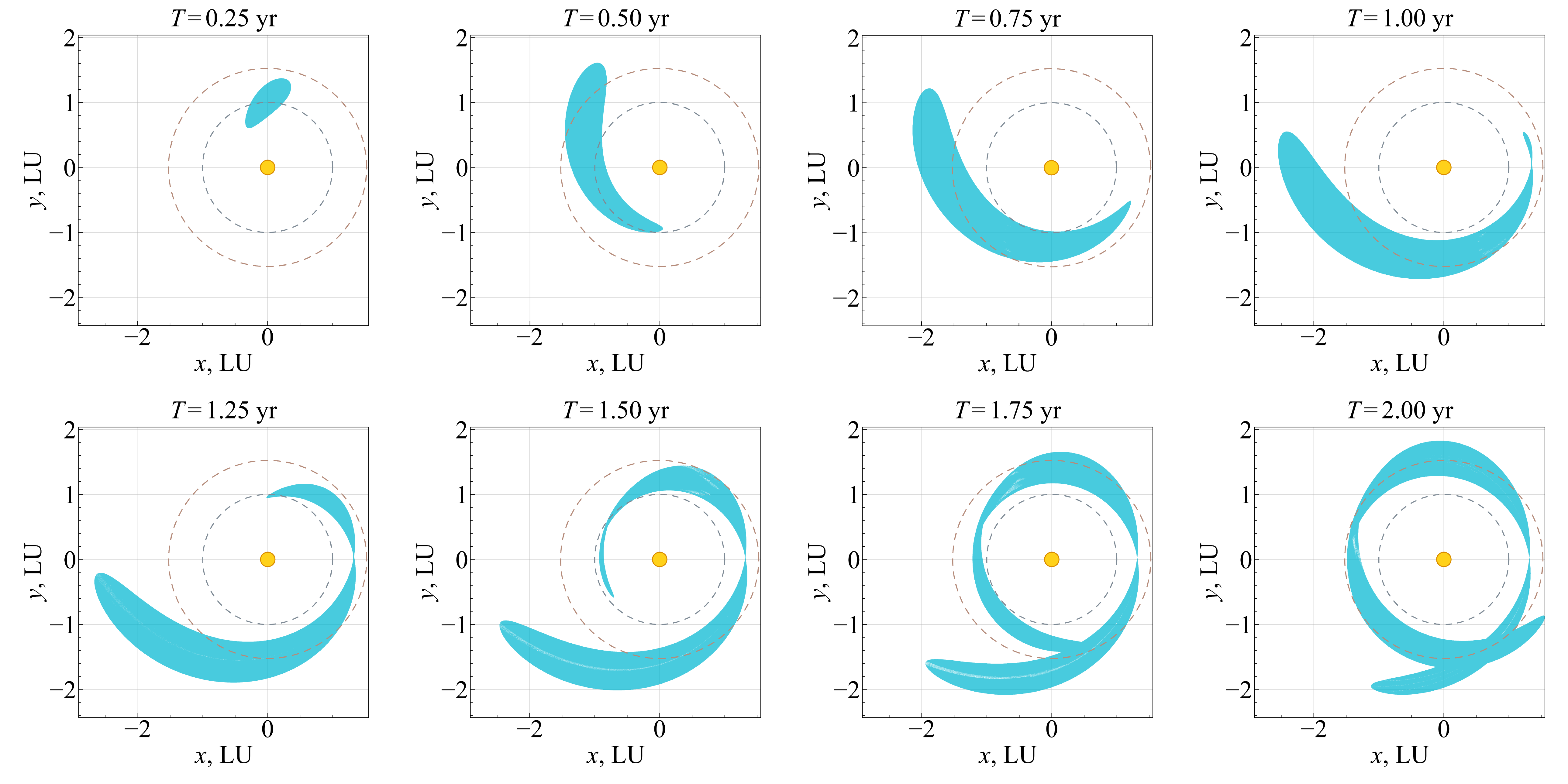}}
\caption{Configurations of reachable sets (TDS, $\theta=\pi/3$)}\label{fig_TDS_60}
\end{figure}

Figures \ref{fig_RDS_30}-\ref{fig_RDS_90} present reachable sets using an RDS under different values of $\theta$ and $T$, while reachable sets using a TDS are presented in Figs. \ref{fig_TDS_30}-\ref{fig_TDS_60}. In these figures, the circle with radius set to 1.524 LU \cite{quarta2022solar} represents the Mars orbit. As shown in these figures, when $T=0.25\text{ yr}$, the configurations of the reachable sets exhibit approximate convexity. However, as $T$ increases, the reachable sets lose convexity and eventually become nonconvex as the nonlinear effects become pronounced. Generally, when $T$ is relatively long (1.00-2.00 yr), reachable sets in all cases form configurations that extend around the Sun, with an overall increase in heliocentric distance. Moreover, as $T$ increases, differences in the configurations of the reachable sets become pronounced due to the accumulation of differences in the SRP acceleration. These configurations of the reachable sets reveal the reachability of a diffractive sail with $v_{{\infty},\text{ }d} \in \left[0,\text{ }5\right]\text{ }\left(\text{km/s}\right)$. From the perspective of interplanetary mission design, the developed reachable sets provide information about the positions in heliocentric space that a diffractive sail can reach within a specific range of capability of launch vehicle engines. Subsequently, a preliminary design of the Earth-Mars transfer trajectories is presented based on the aforementioned reachable sets.

\section{Reachable-Set-Based Preliminary Design of the Earth-Mars Transfers}\label{sec4}
This section presents the reachable-set-based preliminary design of the Earth-Mars transfers. First, we propose a design method that generates initial guesses using reachable sets and trajectory correction. Then, the results are presented and discussed.
\subsection{Design Method}\label{subsec4.1}
Due to the absence of closed-form solutions of Eq. \eqref{eq1}, the transfer trajectories should be designed using a numerical method. As mentioned in Section \ref{sec2}, the design procedure includes generation of initial guesses and trajectory correction. In Section \ref{sec3}, we obtain the diffractive-sail reachable sets under different parameter settings, providing a useful insight into generating initial guesses. Then, we propose the reachable-set-based design method for the Earth-Mars transfers.
\subsubsection{Reachable-Set-Based Generation of Initial Guesses}\label{subsubsec4.1.1}
As shown in Figs. \ref{fig_RDS_30}-\ref{fig_RDS_90}, when $T=0.25 \text{ yr}$, there is no intersection point between the reachable sets and the Mars orbit, for all considered values of $\theta$ and types of diffractive sails. It illustrates that the diffractive sail cannot reach the Mars orbit with $v_{{\infty},\text{ }d} \in \left[0,\text{ }5\right]\text{ }\left(\text{km/s}\right)$ within 0.25 yr. When $T$ is longer than 0.25 yr, the intersection points between the reachable sets and the Mars orbit can be observed, indicating the feasibility of the Earth-Mars transfers within these values of $T$. Then, we use these reachable sets to generate the initial guesses. We focus on the transfers with a relatively short $T$. Therefore, we select the reachable sets with $T=0.50,\text{ }0.75, \text{ and }1.00\text{ yr}$. We further select samples that satisfy $r_f=\sqrt{x_f^2+y_f^2}\in \left[0.98 r_{\text{Mars}},\text{ }1.02r_{\text{Mars}}\right]$ ($r_{\text{Mars}}=1.524\text{ LU}$) in these reachable sets and record the corresponding $\left(\alpha,\text{ }v_{{\infty},\text{ }d}\right)$. Subsequently, trajectory correction is performed to make the trajectories satisfy the specific constraints (i.e., reach the Mars orbit).
\subsubsection{Trajectory Correction}\label{subsubsec4.1.2}
In this paper, trajectory correction is performed by the sequential least-squares programming (SLSQP) algorithm implemented by the scipy.optimize routine in SciPy 1.17.1. The optimization variables are set as $\bm{y}=\left[\alpha,\text{ }v_{\infty,\text{ }d}\right]^{\text{T}}$, and the constraint that the trajectories should satisfy is set as:
\begin{equation}
x_f^2+y_f^2-r_{\text{Mars}}^2=0
\label{eq17}
\end{equation}
and the objective function is set to 0 (i.e., the trajectory correction performed in this paper is formulated as a feasibility problem). Notably, in trajectory correction, $T$ is fixed, and the trajectory state $\left[x_f,\text{ }y_f,\text{ }u_f,\text{ }v_f\right]^\text{T}$ is obtained by the RKF78 integration, not the DA propagation. The parameter setting of the SLSQP algorithm is presented in Table \ref{parameter_setting}.
\begin{table}[!htb]
\caption{Parameter setting of the SLSQP algorithm}\label{parameter_setting}%
\centering
\renewcommand{\arraystretch}{1.5}
\begin{tabular}{@{}ll@{}}
\hline
Parameter & Value  \\
\hline
ftol & $1\times 10^{-8}$ \\
eps & $1\times 10^{-10}$ \\
maxiter & 5000 \\
\hline
\end{tabular}
\end{table}

After the trajectory correction, we select the solutions satisfying $\left|x_f^2+y_f^2-r_{\text{Mars}}^2\right|<1\times 10^{-8}$ as the feasible solutions. To evaluate the performance of the feasible solutions, the departure characteristic energy $C_{3d}$ and arrival characteristic energy $C_{3a}$ are computed as:
\begin{equation}
C_{3d}=v_{\infty,\text{ }d}^2
\label{eq18}
\end{equation}
\begin{equation}
C_{3a}=\left(u_f-u_{\text{Mars}}\right)^2 + \left(v_f-v_{\text{Mars}}\right)^2
\label{eq19}
\end{equation}
where:
\begin{equation}
\begin{gathered}
   {u_{{\text{Mars}}}} =  - {V_{{\text{Mars}}}}\sin \gamma   \hfill \\
   {v_{{\text{Mars}}}} = {V_{{\text{Mars}}}}\cos \gamma    \hfill \\
   {V_{{\text{Mars}}}} = \sqrt {\frac{1}{{{r_{{\text{Mars}}}}}}}    \hfill \\
   \gamma  = {\text{atan2}}\left( {{y_f},{\text{ }}{x_f}} \right) \hfill \\
\end{gathered}
\label{eq20}
\end{equation}
Due to the existence of the Mars atmosphere, the diffractive sail can perform aerobraking when arriving in the vicinity of Mars \cite{yao2022nonsingular}. Therefore, we typically focus on reducing $C_{3d}$. Subsequently, the corresponding design results are presented and analyzed.

\textit{Remark 4}: The proposed method can be applied to the design of other interplanetary transfer scenarios (an example of the Earth-Venus transfer can be found in the Appendix). To achieve this extension, the parameters to compute the reachable sets (e.g., $v_{\infty,\text{ }d}$ and $T$) can be adjusted accordingly.

\textit{Limitation}: This paper focuses on the preliminary design of the Earth-Mars transfers. Therefore, reachability to the Mars orbit is considered and addressed, but the phase relationship between Earth and Mars (i.e., launch window) is not considered. To extend the design results in this paper to a practical interplanetary mission, this relationship should be further added to the design procedure.

\subsection{Results and Discussion}\label{subsec4.2}
\begin{figure}[H]
\centerline{\includegraphics[width=0.96\textwidth]{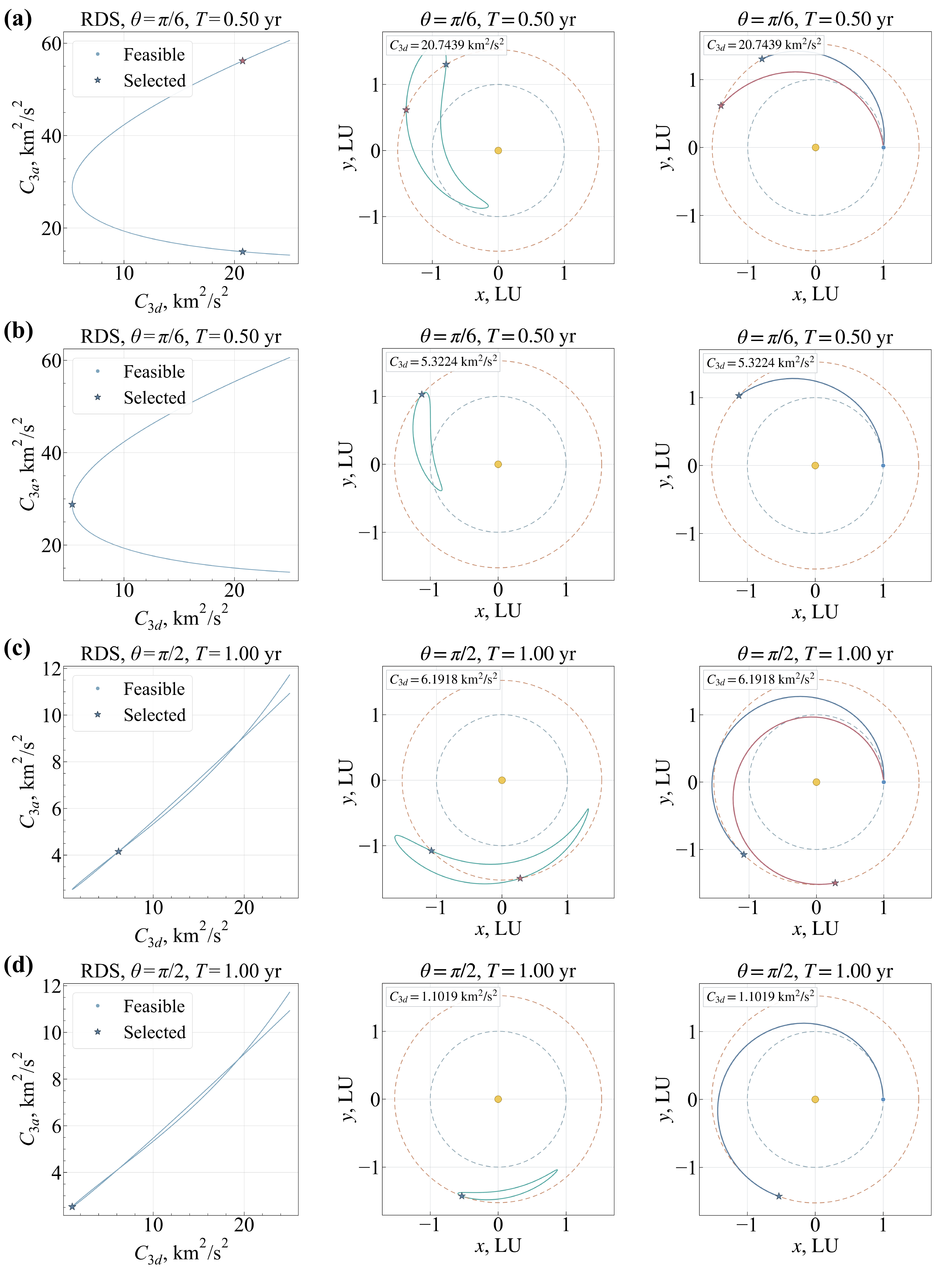}}
\caption{Illustration of the $\left(C_{3d},\text{ }C_{3a}\right)$ distributions. (a) $\theta=\pi/6,\text{ } T=0.50\text{ yr}$, two solutions; (b) $\theta=\pi/6,\text{ } T=0.50\text{ yr}$, single solution; (c) $\theta=\pi/2,\text{ } T=1.00\text{ yr}$, two solutions; (d) $\theta=\pi/2,\text{ } T=1.00\text{ yr}$, single solution.}\label{fig_illustration}
\end{figure}

After trajectory correction, we obtain $\left(C_{3d},\text{ }C_{3a}\right)$ distributions for all cases within one year. Figure \ref{fig_illustration} presents two examples of the $\left(C_{3d},\text{ }C_{3a}\right)$ distributions using an RDS. Within one year ($T=0.5,\text{ }0.75,\text{ }1.00\text{ yr}$), the $\left(C_{3d},\text{ }C_{3a}\right)$ distributions are typically divided into two branches, i.e., generally, there exist two solutions with the same $C_{3d}$. The physical interpretation for this distribution is that the corresponding equal-$v_{\infty,\text{ }d}$ contour (also the equal-$C_{3d}$ contour) has two intersection points with the Mars orbit, yielding two transfer trajectories with the same $C_{3d}$. Moreover, there exists a minimum $C_{3d}$ to allow a feasible solution, which corresponds to the case where the corresponding equal-$v_{\infty,\text{ }d}$ contour is tangent to the Mars orbit. For longer $T$, more branches of feasible solutions can be identified as intersection points with the Mars orbit increase. Notably, when $\theta=\pi/2$, two self-intersection points on the $\left(C_{3d},\text{ }C_{3a}\right)$ distribution can be observed. Each self-intersection point corresponds to two different Earth-Mars transfer trajectories but with the same $C_{3d}$ and $C_{3a}$, whose physical meaning is different from that of the minimum $C_{3d}$ point. Then, a comparison with the obtained solutions in the heliocentric two-body problem is performed to highlight the advantage of using a diffractive sail. 

\begin{figure}[H]
\centerline{\includegraphics[width=0.96\textwidth]{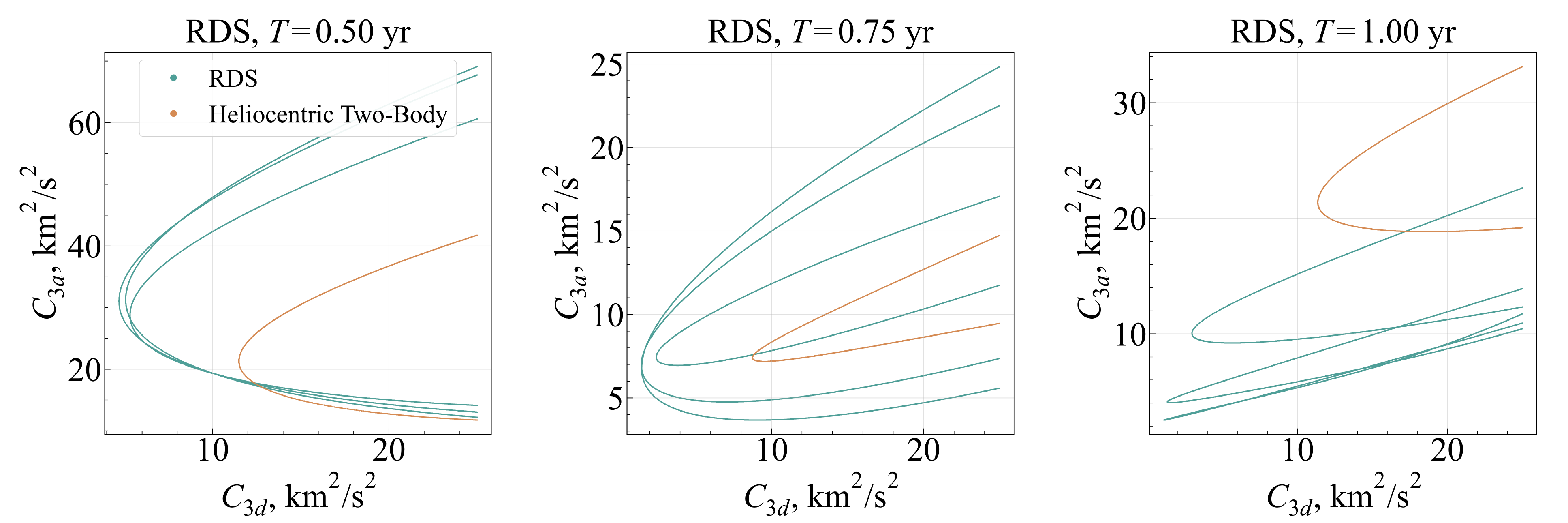}}
\caption{Comparison in terms of the $\left(C_{3d},\text{ }C_{3a}\right)$ distributions (RDS).}\label{fig_comparison_RDS}
\end{figure}

\begin{figure}[H]
\centerline{\includegraphics[width=0.96\textwidth]{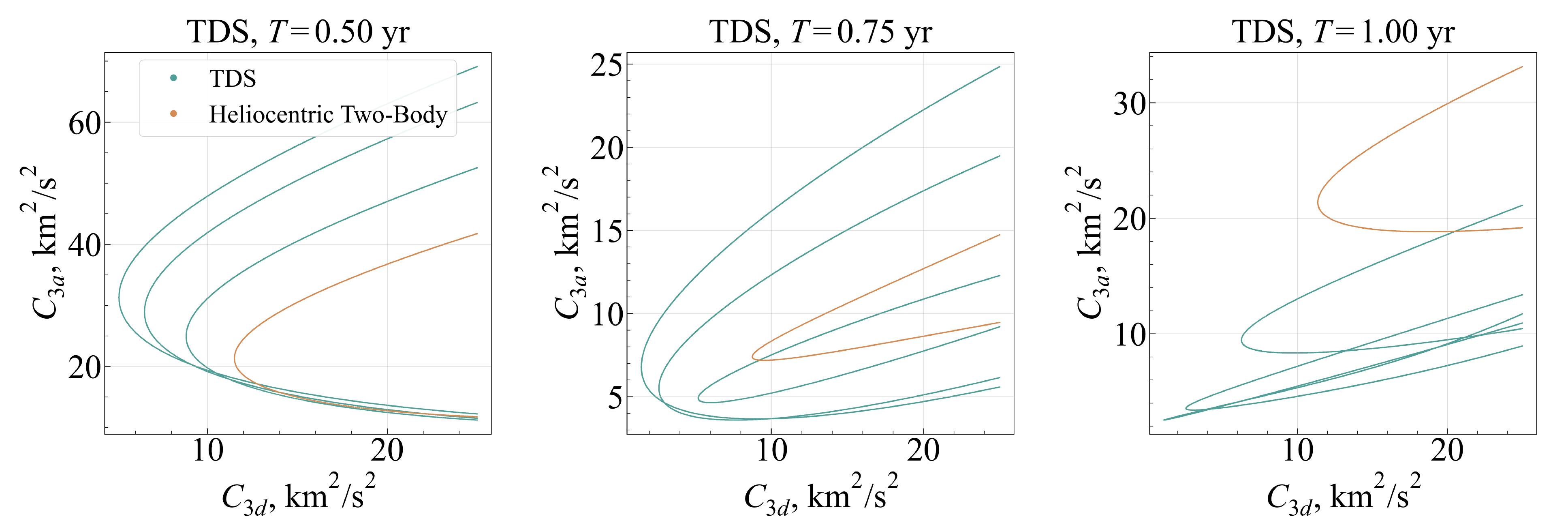}}
\caption{Comparison in terms of the $\left(C_{3d},\text{ }C_{3a}\right)$ distributions (TDS).}\label{fig_comparison_TDS}
\end{figure}

Figures \ref{fig_comparison_RDS}-\ref{fig_comparison_TDS} present the $\left(C_{3d},\text{ }C_{3a}\right)$ distributions using an RDS/TDS and the heliocentric two-body problem. The $\left(C_{3d},\text{ }C_{3a}\right)$ distributions in the heliocentric two-body problem are obtained from the same procedure mentioned in Section \ref{sec3} and Section \ref{subsec4.1}, including the reachable-set computation using DA, reachable-set-based generation of initial guesses, and trajectory correction, but in the heliocentric two-body problem (i.e., $\beta=0$). For each subfigure, three $\left(C_{3d},\text{ }C_{3a}\right)$ distributions using an RDS/TDS denote $\left(C_{3d},\text{ }C_{3a}\right)$ distributions with $\theta=\pi/6,\text{ }\pi/3,\text{ and }\pi/2$, respectively. From these figures, it can be observed that using a diffractive sail can effectively reduce the required minimum $C_{3d}$ of transfers with a fixed $T$. Moreover, when $T=0.75\text{ and }1.00\text{ yr}$, using diffractive sails under specific values of $\theta$ can achieve a reduction in both $C_{3d}$ and $C_{3a}$. Table \ref{table_comparison} presents the comparison results in terms of the minimum $C_{3d}$. It can be found that when $T=0.50\text{ yr}$, using an RDS can achieve up to 59.13$\%$ of reduction in the minimum $C_{3d}$, while using a TDS can achieve up to 43.48$\%$ of reduction; when $T=0.75\text{ yr}$, using an RDS can achieve up to 83.43$\%$ of reduction, while using a TDS can achieve up to 70.06$\%$ of reduction; when $T=1.00\text{ yr}$, using an RDS can achieve up to 90.32$\%$ of reduction, while using a TDS can achieve up to 77.46$\%$ of reduction. As $T$ increases, the effects of using a diffractive sail become pronounced. These comparison results reveal the advantage of using a diffractive sail in the interplanetary mission. Subsequently, the effects of the type of the diffractive sail, $\theta$, and $T$ on the $\left(C_{3d},\text{ }C_{3a}\right)$ distributions are analyzed, which can provide some insights into the parameter selection for the diffractive-sail interplanetary mission design.

\begin{table}[!htb]
\caption{Comparison in terms of the minimum $C_{3d}$}\label{table_comparison}%
\centering
\renewcommand{\arraystretch}{1.5}
\begin{tabular}{@{}llll@{}}
\hline
Type & $\theta$ & $T,\text{ yr}$ & min $C_{3d},\text{ }\text{km}^2/\text{s}^2$ \\
\hline
Heliocentric Two-Body & -- & 0.50 & 11.50 \\
Heliocentric Two-Body & -- & 0.75 & 8.75 \\
Heliocentric Two-Body & -- & 1.00 & 11.36 \\
RDS  & $\pi/6$    & 0.50    & 5.32     \\
RDS  & $\pi/6$    & 0.75    & 2.42     \\
RDS  & $\pi/6$    & 1.00    & 2.96     \\
RDS  & $\pi/3$    & 0.50    & 4.70     \\
RDS  & $\pi/3$    & 0.75    & 1.45     \\
RDS  & $\pi/3$    & 1.00    & 1.32     \\
RDS  & $\pi/2$    & 0.50    & 5.08     \\
RDS  & $\pi/2$    & 0.75    & 1.47     \\
RDS  & $\pi/2$    & 1.00    & 1.10     \\
TDS  & $\pi/6$    & 0.50    & 8.82     \\
TDS  & $\pi/6$    & 0.75    & 5.20    \\
TDS  & $\pi/6$    & 1.00    & 6.27     \\
TDS  & $\pi/3$    & 0.50    & 6.50     \\
TDS  & $\pi/3$    & 0.75    & 2.62     \\
TDS  & $\pi/3$    & 1.00    & 2.56     \\

\hline
\end{tabular}
\end{table}

\begin{figure}[H]
\centerline{\includegraphics[width=0.96\textwidth]{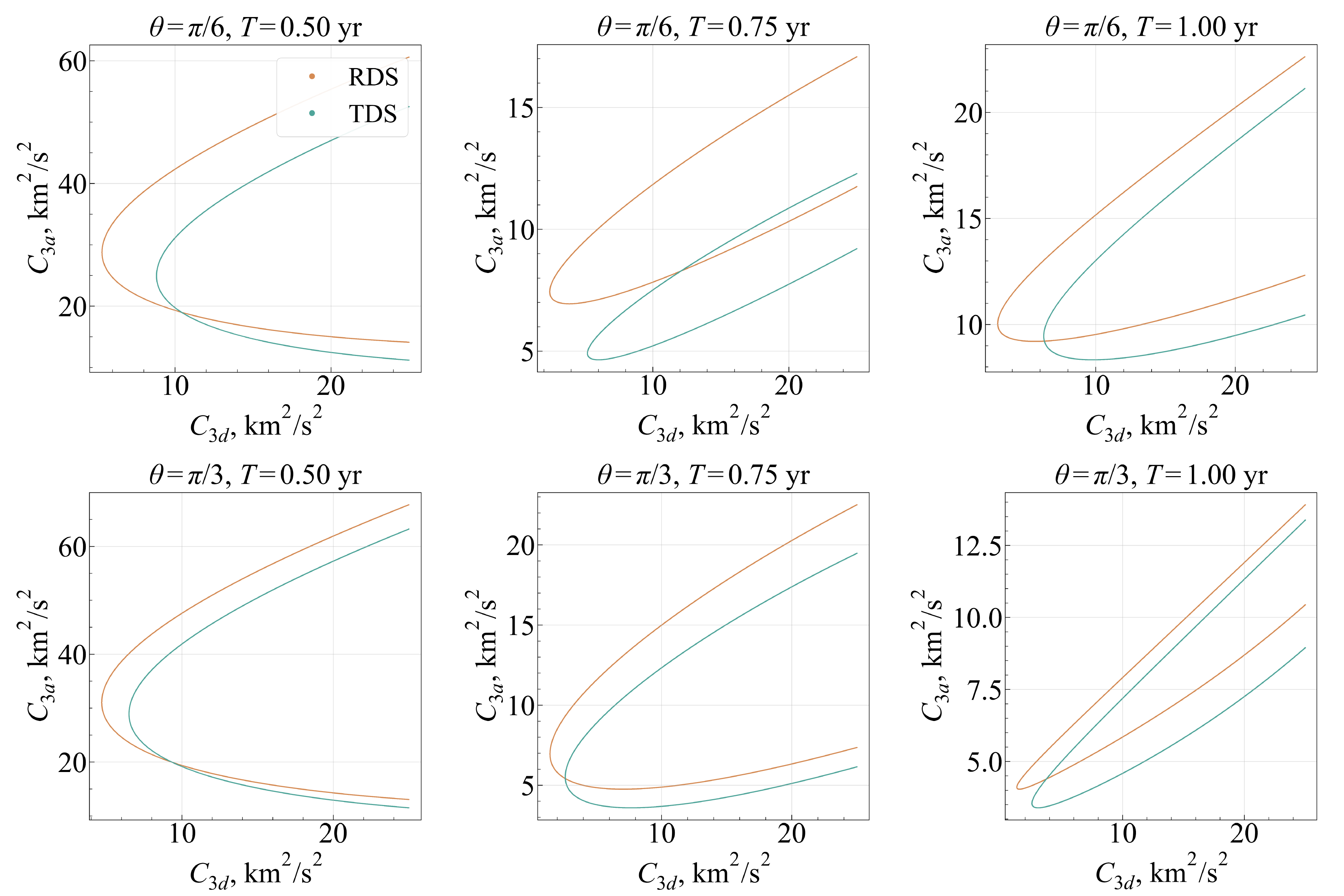}}
\caption{The $\left(C_{3d},\text{ }C_{3a}\right)$ distributions for different types of diffractive sails.}\label{fig_C3d_C3a_RDS_TDS}
\end{figure}

Figure \ref{fig_C3d_C3a_RDS_TDS} presents the comparison results of using RDS and TDS (when $\theta=\pi/2$, RDS is equivalent to TDS). It can be observed that for all cases, the minimum $C_{3d}$ using an RDS is lower than that using a TDS. Therefore, \textit{for the Earth-Mars transfers, it is suggested to use an RDS rather than a TDS}. The subsequent analysis focuses on the results using an RDS.

\begin{figure}[H]
\centerline{\includegraphics[width=0.96\textwidth]{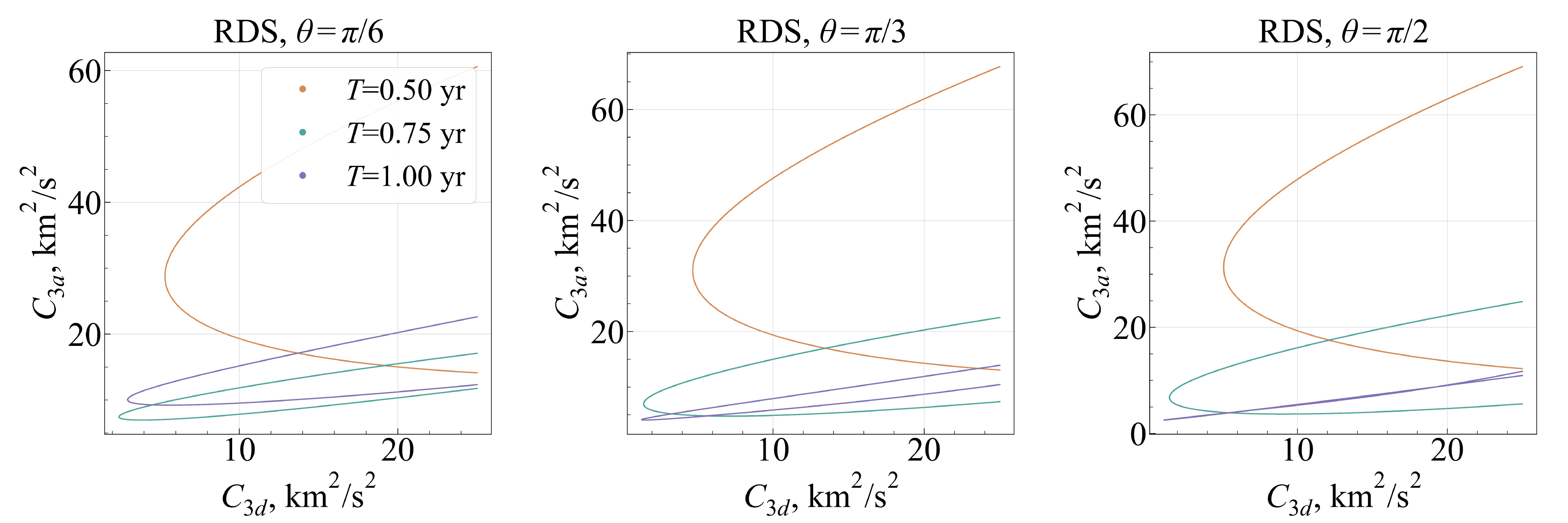}}
\caption{The $\left(C_{3d},\text{ }C_{3a}\right)$ distributions under the fixed $\theta$.}\label{fig_C3d_C3a_fixedtheta}
\end{figure}

Figure \ref{fig_C3d_C3a_fixedtheta} presents the $\left(C_{3d},\text{ }C_{3a}\right)$ distributions under the fixed $\theta$ to reveal the effects of $T$ (RDS). From Fig. \ref{fig_C3d_C3a_fixedtheta}, it can be observed that when $T$ is longer than 0.50 yr (i.e., $T=0.75\text{ and }1.00\text{ yr}$), the obtained values of minimum $C_{3d}$ are comparable, with a relatively pronounced reduction in $C_{3d}$ compared with the minimum $C_{3d}$ solution under $T=0.50\text{ yr}$. Based on the aforementioned discussion, from the perspective of mission design, \textit{it is suggested that selecting a transfer time longer than 0.50 yr effectively reduces the required $C_{3d}$}. Then, we select the obtained solutions with $T$ longer than 0.50 yr and analyze the effects of $\theta$.

\begin{figure}[H]
\centerline{\includegraphics[width=0.64\textwidth]{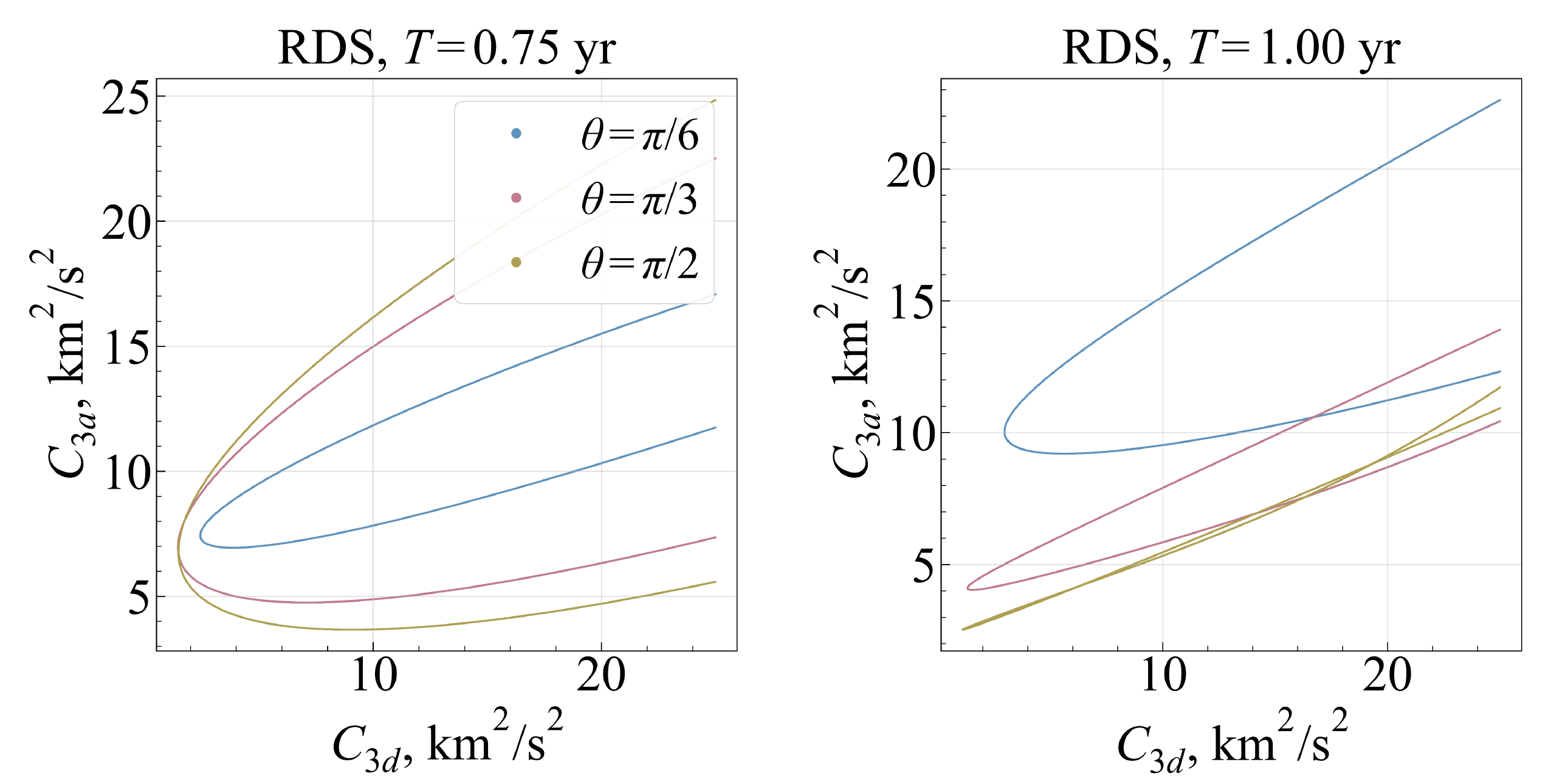}}
\caption{The $\left(C_{3d},\text{ }C_{3a}\right)$ distributions under the fixed $T$.}\label{fig_C3d_C3a_fixedtof}
\end{figure}

\begin{figure}[H]
\centerline{\includegraphics[width=0.64\textwidth]{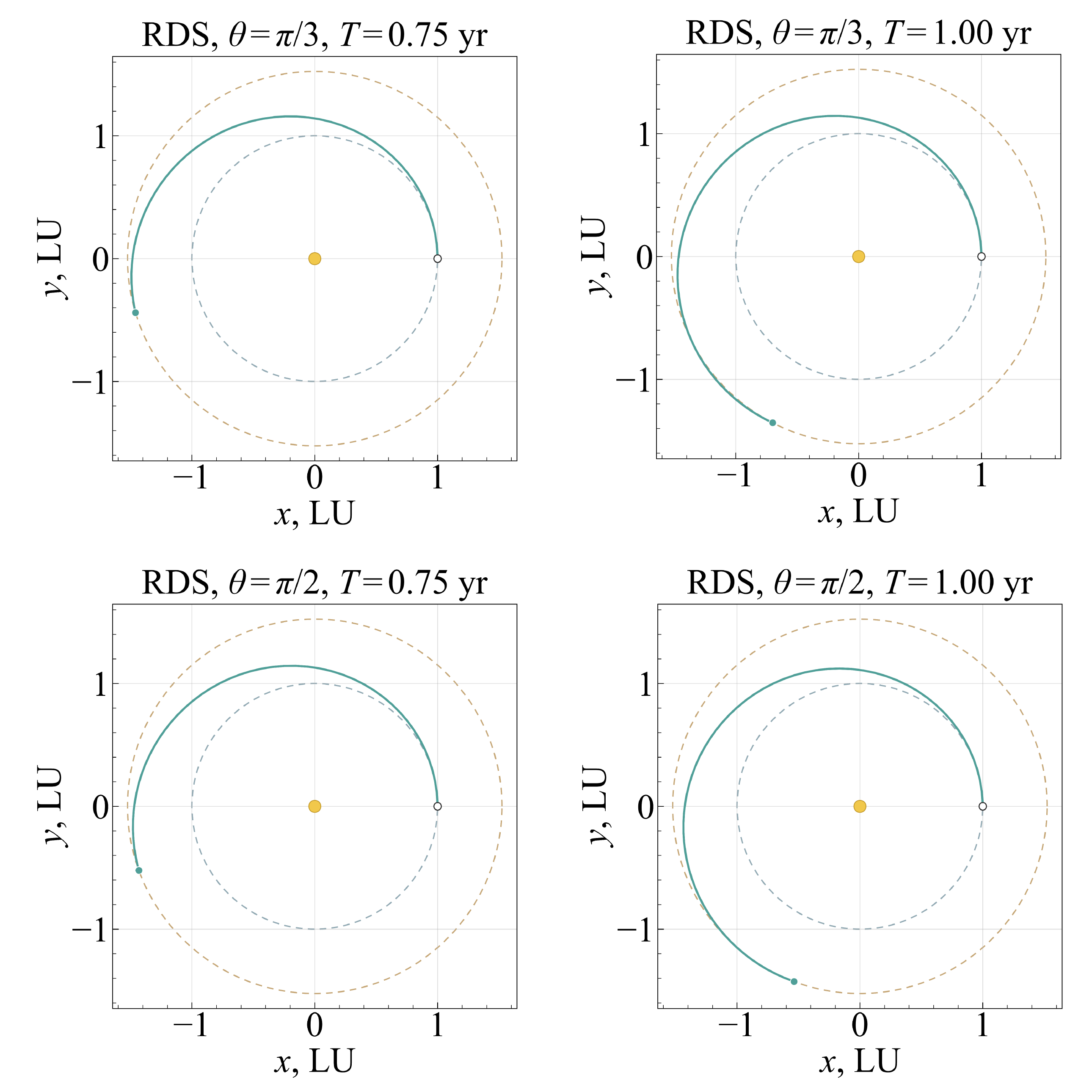}}
\caption{Typical trajectories with the minimum $C_{3d}$ (RDS).}\label{fig_trajectory_RDS}
\end{figure}

Figure \ref{fig_C3d_C3a_fixedtof} presents the $\left(C_{3d},\text{ }C_{3a}\right)$ distributions under the fixed $T$ (0.75 yr and 1.00 yr) to reveal the effects of $\theta$. It can be observed that the cases under $\theta=\pi/3$ and $\theta=\pi/2$ yield comparable minimum $C_{3d}$, whose values are lower than those under $\theta=\pi/6$. Therefore, \textit{for the Earth-Mars transfer, it is suggested to select an RDS under $\theta=\pi/3$ or $\theta=\pi/2$, not under $\theta=\pi/6$}. Notably, this conclusion is drawn within a specific range of $\theta$ (i.e, $\theta=\pi/6,\text{ }\pi/3,\text{ and }\pi/2$); more values of $\theta$ yielding lower values of the minimum $C_{3d}$ can be identified by performing a corresponding analysis on a larger set of $\theta$ values. Then, typical trajectory samples (under $\theta=\pi/3, \text{ }\pi/2$ and $T=0.75,\text{ }1.00\text{ yr}$) are presented in Fig. \ref{fig_trajectory_RDS}. The corresponding values of $C_{3d}$ and $C_{3a}$ are presented in Table \ref{table_trajectory_RDS}. It is observed that for these four cases, the increase of $T$ causes the reduction of the minimum $C_{3d}$ and the corresponding $C_{3a}$, under both $\theta=\pi/3$ and $\theta=\pi/2$. Based on the aforementioned discussion, the reachable sets provide useful information about the reachability of a diffractive sail under specific parameters and generate initial guesses for the interplanetary transfers. For the application of the designed transfers to the practical mission, the launch window \cite{battin1999introduction}, escape from Earth \cite{mengali2004earth,macdonald2005realistic,terzaghi2026performance,fu2026energy}, and capture with respect to Mars \cite{hyeraci2010method,dei2018survey,yao2022nonsingular} should be further considered.

\begin{table}[!htb]
\caption{Transfer characteristics of typical trajectories (RDS)}\label{table_trajectory_RDS}%
\centering
\renewcommand{\arraystretch}{1.5}
\begin{tabular}{@{}llll@{}}
\hline
$\theta$ & $T,\text{ yr}$  & $C_{3d},\text{ }\text{km}^2/\text{s}^2$ & $C_{3a},\text{ }\text{km}^2/\text{s}^2$\\
\hline
$\pi/3$    & 0.75    & 1.45 & 6.96    \\
$\pi/3$    & 1.00    & 1.32 & 4.11    \\
$\pi/2$    & 0.75    & 1.47 & 6.79    \\
$\pi/2$    & 1.00    & 1.10 & 2.53    \\
\hline
\end{tabular}
\end{table}

\section{Conclusion}\label{sec5}
This paper proposes a complementary computational method for reachable sets under arbitrary dynamics and applies it to diffractive-sail reachable sets and interplanetary transfer. The diffractive-sail interplanetary transfer problem is first transformed into the problem of computing single-impulse reachable sets. Here, the impulse represents the departure hyperbolic excess velocity. Then, a computational method for single-impulse reachable sets is proposed using differential algebra combined with adaptive grid refinement. In particular, the adaptive grid refinement is achieved by the quadtree method considering two types of merit scores, one to limit the worst-case local resolution of the reachable set, and the other to limit the maximum truncation error of the differential algebra propagation. The results of the reachable-set computation and global evaluation verify the effectiveness of the proposed method. The configurations of the obtained diffractive-sail reachable sets are presented and analyzed, providing information about the final positions in the heliocentric space. Finally, a preliminary design of the Earth-Mars transfers is performed based on the obtained reachable sets, which are used to generate initial guesses. The design results are analyzed and discussed. It is found that compared with the two-body transfers, the introduction of the diffractive sail can reduce the minimum departure characteristic energy within the fixed transfer time by up to 90.32$\%$. Analyzing the effects of the corresponding parameters (transfer time, diffractive angle, type of diffractive sail) on the transfer characteristics, in the considered cases in this paper, it is suggested that selecting the reflection-type diffractive sail under diffractive angles set to $\pi/3$ and $\pi/2$ within a transfer time longer than 0.5 years. Through this work, a direct link between interplanetary transfer, the reachable set, and diffractive-sail dynamics can be established.

\section*{Appendix: An Example of Preliminary Design of the Earth-Venus Transfers}
\begin{figure}[H]
\centerline{\includegraphics[width=0.6\textwidth]{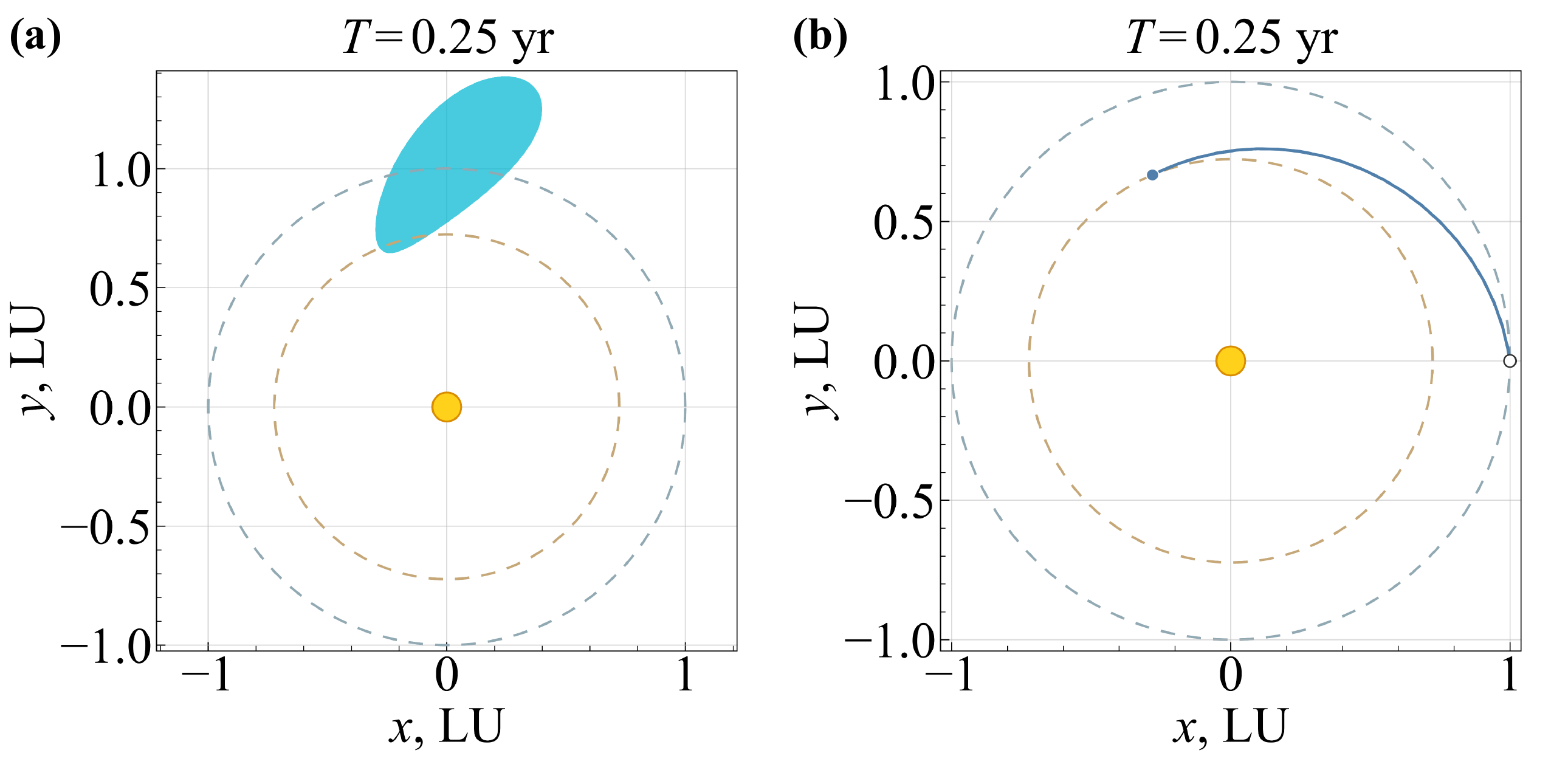}}
\caption{An example of the reachable-set-based construction of the Earth-Venus transfer trajectory. (a) Reachable set; (b) Earth-Venus transfer trajectory.}\label{fig_venus}
\end{figure}

Figure \ref{fig_venus} presents an example of preliminary design of the Earth-Venus transfers. As shown in Fig. \ref{fig_venus} (a), the reachable set using an RDS under $\theta=\pi/6$ and $T=0.25\text{ yr}$ has intersection points with the Venus orbit (whose radius is set to 0.723 LU \cite{quarta2022solar}), which yields the Earth-Venus transfer trajectories. A trajectory sample is presented in Fig. \ref{fig_venus} (b).

\section*{Acknowledgments}
This work was supported by the National Natural Science Foundation of China (Grant Nos. 12372044, 12525204, and U23B6002).

\bibliography{sample}

\end{document}